\documentclass{article}

\usepackage[T1]{fontenc}
\usepackage[utf8]{inputenc}
\usepackage{hyperref}
\usepackage{url}
\usepackage{amsmath}
\usepackage{amssymb}
\usepackage{bm}
\usepackage{graphicx}
\usepackage{subcaption}
\usepackage{booktabs}
\usepackage{multirow}
\usepackage{xcolor}
\usepackage{enumitem}
\usepackage{microtype}
\usepackage{comment}
\usepackage[accepted]{mlsys2025}

\makeatletter
\renewcommand{\Notice@String}{}
\makeatother

\hypersetup{
  pdfsubject={LLM serving, scheduling, and long-context inference}
}

\usepackage{xr-hyper}
\usepackage{float}
\usepackage{placeins}


\newcommand{\maybeincludegraphics}[2][]{%
  \IfFileExists{#2}{\includegraphics[#1]{#2}}{%
    \fbox{\parbox{0.88\linewidth}{\centering Missing figure: \texttt{\detokenize{#2}}}}%
  }%
}

\mlsystitlerunning{P-PAS: Prefill-Pressure Adaptive Scheduling}

\begin{document}

\twocolumn[
\mlsystitle{P-PAS: Prefill-Pressure Adaptive Scheduling for Long-Context LLM Serving}

\begin{mlsysauthorlist}
\mlsysauthor{Timo S{\"a}mann}{independent}
\end{mlsysauthorlist}
\mlsysaffiliation{independent}{Independent Researcher, Germany}
\mlsyscorrespondingauthor{Timo S{\"a}mann}{timo.saemann@gmx.de}
\mlsyskeywords{LLM serving, scheduling, chunked prefill, long-context inference, vLLM}
\vskip 0.3in
]
\printAffiliationsAndNotice{}

\begin{abstract}
Long-context LLM applications such as retrieval-augmented generation (RAG)
and agentic systems often process tens of thousands of input tokens to produce
short outputs, making end-to-end request latency an important serving
objective. We show that the maximum number of batched tokens (MBT), which
controls the token scheduling budget in vLLM, has a scheduling-pressure-dependent effect on latency. Larger token budgets can reduce latency under low
scheduling pressure, while smaller budgets become preferable under higher
pressure. Consequently, no single static MBT performs best across load
regimes.

We introduce Prefill-Pressure Adaptive Scheduling (P-PAS), a lightweight
policy that dynamically adapts the global scheduling budget using concurrent
prefill and decode state. P-PAS retains a large token budget under low pressure
and switches to a smaller budget when multiple prefills compete with active
decoding. Across models, workload configurations, and GPUs, P-PAS maintains
low end-to-end latency across changing load regimes. Kernel-level profiling
shows that the crossover arises from a trade-off between large-chunk prefill
efficiency and prefill--decode interference.

We further extend the pressure signal to heterogeneous request shapes with
P-PAS+. On mixed-shape workloads using Qwen3.8-27B and
Nemotron-3.5-Lightning-30B-A3B, P-PAS+ reduces mean end-to-end latency by
3.0--3.1\% relative to the best tested static token budget. Across exactly
matched logical requests, 91.2\% of Qwen requests and 88.5\% of Nemotron
requests complete faster with P-PAS+. Code and artifacts are available in the
\href{https://github.com/TimoSaemann/ppas-vllm}{public repository}.
\end{abstract}

\section{Introduction}

\begin{figure}[t]
    \centering
    \includegraphics[width=1.0\linewidth]{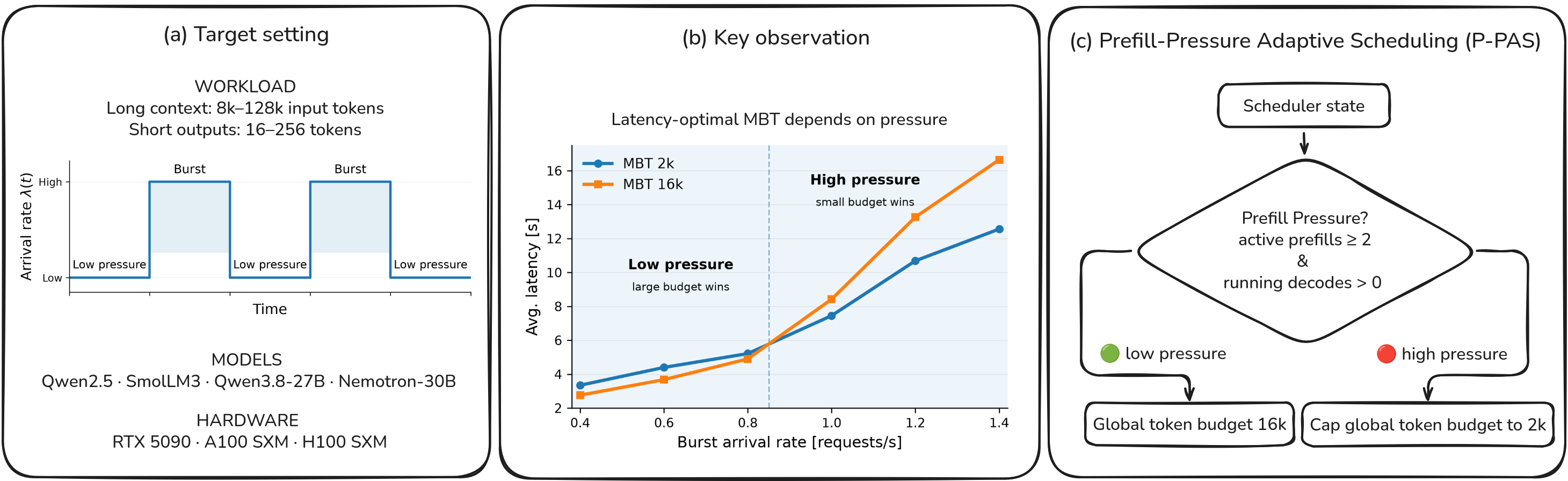}
    \caption{Motivation and overview of P-PAS.
    (a) (a) We target long-context, short-output serving under changing and
bursty load, including heterogeneous request shapes, across multiple
model scales and GPU classes.
    (b) The latency-optimal maximum batched tokens (MBT) depends on scheduling
    pressure: large budgets can perform better under low pressure, while
    smaller budgets become preferable as pressure increases.
    (c) P-PAS switches between two calibrated scheduling-budget endpoints
    using prefill pressure and active decode state.}
    \label{fig:overview}
\end{figure}

Large language model (LLM) serving increasingly extends beyond interactive
chat toward applications that consume large amounts of context while producing
comparatively short outputs. Retrieval-augmented generation (RAG)~\cite{lewis2020retrieval},
document analysis, and agentic systems can provide an LLM with retrieved
documents, tool outputs, execution traces, or long conversation histories while
requiring only a short answer or action in return~\cite{she2026laps}. We focus
on such long-context, short-output workloads under bursty request arrivals, as
illustrated in Figure~\ref{fig:overview}(a). In agentic workflows in particular,
downstream actions often depend on the completed LLM response~\cite{yao2023react},
motivating our focus on end-to-end request latency.

Serving engines for autoregressive LLMs interleave prefill and decode work on
the GPU, making their interaction important for end-to-end
latency~\cite{agrawal2024taming}. vLLM~\cite{kwon2023efficient}, a widely used
LLM serving engine, uses continuous batching and chunked prefill to manage
concurrent requests. A key scheduling parameter is the maximum number of
batched tokens (MBT), which limits the token budget available within one
scheduler iteration. Larger MBTs allow more prefill tokens to be processed
together, whereas smaller MBTs divide long prefills into smaller chunks.

We show that the latency-optimal MBT depends on scheduling pressure. Under low
pressure, larger token budgets can reduce end-to-end latency through more
efficient prefill execution, while under high pressure, smaller budgets become
preferable by limiting interference with active decoding. Consequently, no
single static MBT performs best across load regimes. Figure~\ref{fig:overview}(b)
illustrates this behavior and motivates adaptive control of the scheduling
budget.

To adapt to this load-dependent behavior, we introduce
\emph{Prefill-Pressure Adaptive Scheduling} (P-PAS), a lightweight extension
to the vLLM scheduler. P-PAS dynamically adapts the global scheduling budget
based on prefill pressure and decode state. It retains a large budget under low
pressure and constrains the budget when concurrent prefills compete with active
decoding. P-PAS operates directly on scheduler state and requires neither
offline workload prediction nor a predefined request-rate threshold.

We evaluate P-PAS primarily using Qwen2.5-3B on an NVIDIA RTX 5090 with
25k-token prompts and 32-token outputs, and further assess its robustness
across model architectures and scales, prompt lengths from 20k to 30k tokens,
output lengths from 16 to 64 tokens, and consumer- and datacenter-class GPUs.
Across these configurations, P-PAS maintains low end-to-end latency as
scheduling pressure changes, avoiding the need to select a single fixed MBT
for different load regimes.

We then extend the study to heterogeneous request shapes and substantially
larger recent models. Request count alone can understate prefill pressure when
a single request has a very large amount of remaining prefill work. P-PAS+
augments the original count-based pressure signal with this residual-work
information. On heterogeneous Qwen3.8-27B and
Nemotron-3.5-Lightning-30B-A3B workloads, P-PAS+ improves mean end-to-end
latency by 3.0--3.1\% over the best tested static budget while accelerating
the large majority of exactly matched requests.

Finally, we use GPU profiling to investigate the source of MBT sensitivity.
For Qwen2.5, larger prefill chunks can reduce the cumulative execution time of
dominant attention and MLP kernels compared with processing the same prompt
through multiple smaller chunks. This prefill-efficiency advantage explains
why large MBTs can reduce latency under low pressure. As pressure increases,
however, larger prefill chunks interfere more strongly with active decoding,
shifting the latency-optimal configuration toward smaller budgets.

In summary, this work makes the following contributions:
\begin{itemize}[leftmargin=*]
    \item We characterize \textbf{MBT sensitivity} in long-context,
    short-output LLM serving and identify a load-dependent crossover in which
    the latency-optimal token budget changes with scheduling pressure.

    \item We introduce \textbf{P-PAS}, a lightweight scheduling policy that
    dynamically adapts the global token budget based on prefill pressure and
    active decode state, avoiding a single static MBT choice.

    \item We provide a simple \textbf{endpoint-calibration view} that separates
    the question of which two budgets represent the low- and high-pressure
    operating points from the online decision of when to switch between them.

    \item We evaluate P-PAS across \textbf{model architectures and scales,
    workload configurations, and consumer- and datacenter-class GPUs}, and
    provide a \textbf{kernel-level analysis} of the underlying crossover.

    \item We introduce \textbf{P-PAS+} for highly heterogeneous request shapes
    and evaluate it on Qwen3.8-27B and Nemotron-30B, where it outperforms the
    best tested static budget on mixed long-context workloads.
\end{itemize}

\section{Background and Motivation: MBT Sensitivity}
\label{sec:mbt_sensitivity}

Autoregressive LLM inference consists of two phases. During prefill, the model
processes the input tokens and stores the resulting keys and values in the KV
cache. During \emph{decode}, output tokens are generated autoregressively. In
vLLM, continuous batching allows prefill and decode work from different
requests to be scheduled together within an iteration. For long prompts,
chunked prefill divides the prefill into smaller chunks that can be spread
across multiple iterations.

The maximum number of batched tokens (MBT) limits the total number of tokens
that can be scheduled in one iteration. This budget is shared by prefill and
decode work. A large MBT allows more prefill tokens to be processed at once,
reducing the number of chunks required for long prompts, whereas a small MBT
divides the prefill across more scheduling iterations.

This creates a trade-off for long-context workloads. Larger token budgets can
process long prefills efficiently, whereas smaller budgets provide more
frequent opportunities for active decode requests to make progress. The
latency-optimal MBT therefore depends on the balance between prefill efficiency
and decode interference.

Figure~\ref{fig:overview}(b) demonstrates this behavior for Qwen2.5-3B on an
RTX 5090 with 25k-token prompts and 32-token outputs. At low arrival rates,
MBT 16k achieves lower end-to-end latency than MBT 2k. As the arrival rate
increases, the ordering reverses: MBT 2k becomes preferable while the latency
of MBT 16k increases more rapidly. We refer to this dependence of request
latency on the scheduling budget as \emph{MBT sensitivity}.

\paragraph{An idealized view.}
Let \(b\) denote a fixed scheduler token budget, and let
\(L_{\mathrm{low}}(b)\) and \(L_{\mathrm{high}}(b)\) denote the expected
request latency under low- and high-pressure conditions, respectively.
We use \(p\in[0,1]\) as an idealized measure of how strongly a workload is
exposed to high scheduling pressure. The latency of a static budget can then
be approximated as
\begin{equation}
L_{\mathrm{static}}(b,p)
\approx
(1-p)L_{\mathrm{low}}(b) + pL_{\mathrm{high}}(b).
\label{eq:static_mixture}
\end{equation}
The optimal fixed budget is therefore
\begin{equation}
b^*(p)=\arg\min_b L_{\mathrm{static}}(b,p),
\label{eq:static_optimum}
\end{equation}
and can change as the workload shifts between low- and high-pressure
conditions. P-PAS instead tries to approximate
\begin{equation}
L_{\mathrm{P\text{-}PAS}}(p)
\approx
(1-p)L_{\mathrm{low}}(B_{\max})
+
pL_{\mathrm{high}}(B_{\mathrm{cap}}),
\label{eq:ppas_intuition}
\end{equation}
without knowing \(p\) or the future arrival process in advance. This is an
intuition rather than a queueing model: individual requests can span both
regimes and interact through batching, queueing, and KV-cache state, so the
two terms are not strictly independent.

Importantly, pressure is endogenous to the deployment. The same external
arrival rate can create substantial prefill--decode contention on one
model--hardware configuration and remain low-pressure on another. Arrival rate
alone therefore does not define the regime. Bursty workloads can also move
between low- and high-pressure states during a single serving run, making a
single static choice suboptimal for part of the workload.

\section{Prefill-Pressure Adaptive Scheduling (P-PAS)}
\label{sec:P-PAS}

The MBT sensitivity observed in Section~\ref{sec:mbt_sensitivity} suggests
that the scheduling budget should depend on the current execution state.
P-PAS follows a simple principle: retain a large token budget when prefill
pressure is low, but switch to a smaller budget when multiple prefills compete
with active decoding. This preserves large-chunk prefill efficiency when
possible while limiting interference with active decoding as pressure
increases.

P-PAS derives its decision directly from scheduler state. At each scheduling
iteration, it considers running and waiting prefills together with running
decode requests. Let $N_p$ denote the number of running and waiting prefills
and $N_d$ the number of running decode requests. P-PAS detects pressure when
$N_p$ reaches a threshold $N_{\mathrm{th}}$ while at least one decode request
is active:
\begin{equation}
P_{\mathrm{count}}
=
(N_p \ge N_{\mathrm{th}}) \land (N_d>0).
\label{eq:count_pressure}
\end{equation}

Under pressure, P-PAS uses the smaller scheduler token budget
$B_{\mathrm{cap}}$; otherwise, it uses the low-pressure budget $B_{\max}$.
The decision is reevaluated every iteration, so the scheduler immediately
returns to the larger budget when pressure subsides.

We examine the restriction to two budget endpoints in
Appendix~\ref{sec:two_endpoint_analysis}. For the short-output regime targeted
by P-PAS, intermediate MBTs provide little additional latency benefit, while
longer decode-heavy workloads expose a limitation of the binary controller.

Algorithm~\ref{alg:P-PAS} summarizes the resulting scheduling policy.

\begin{algorithm}[t]
\caption{Prefill-Pressure Adaptive Scheduling (P-PAS)}
\label{alg:P-PAS}
\begin{algorithmic}[1]
\STATE $N_p \gets N_{\mathrm{prefill}}^{\mathrm{running}}
                 + N_{\mathrm{prefill}}^{\mathrm{waiting}}$
\STATE $N_d \gets$ number of running decode requests
\IF{$N_p \geq N_{\mathrm{th}} \land N_d > 0$}
    \STATE $B \gets B_{\mathrm{cap}}$
\ELSE
    \STATE $B \gets B_{\max}$
\ENDIF
\STATE Schedule the iteration with global token budget $B$
\end{algorithmic}
\end{algorithm}

P-PAS uses only state already available to vLLM and requires no additional
model execution, workload prediction, or request-rate estimator.

\subsection{Selecting the Budget Endpoints}
\label{sec:calibration}

P-PAS separates two questions that are easy to conflate: \emph{which} token
budgets represent useful low- and high-pressure operating points for a
deployment, and \emph{when} the scheduler should switch between them. The
pressure detector answers the second question online. The endpoint pair can be
selected with a small offline static sweep over a candidate set $\mathcal{B}$.
A simple calibration rule is
\begin{equation}
B_{\max}=\arg\min_{b\in\mathcal{B}}L_{\mathrm{low}}(b),
\qquad
B_{\mathrm{cap}}=\arg\min_{b\in\mathcal{B}}L_{\mathrm{high}}(b).
\label{eq:endpoint_calibration}
\end{equation}
The calibration traces need only expose representative low- and high-pressure
states; P-PAS is subsequently evaluated on dynamic traces that alternate
between them.

For the primary Qwen2.5-3B study, the static MBT sweep in
Section~\ref{sec:evaluation} identifies 16k and 2k as representative endpoints,
so we use $B_{\max}=16$k and $B_{\mathrm{cap}}=2$k with
$N_{\mathrm{th}}=2$. To test transfer rather than retune each experiment, the robustness study in Section~5.2 deliberately keeps this same P-PAS configuration across
additional models, prompt/output lengths, and GPUs. In
Section~\ref{sec:ppas_plus}, where we move to much larger models and highly
heterogeneous request shapes, we again perform a small deployment-specific
static sweep and use endpoints appropriate to each serving configuration.

\section{Experimental Setup}
\label{sec:setup}

\subsection{Workload}

We evaluate long-context, short-output serving workloads representative of
RAG and agentic applications. Unless stated otherwise, requests contain
25k prompt tokens and generate 32 output tokens.

\begin{figure}[t]
    \centering
    \includegraphics[width=1.0\linewidth]{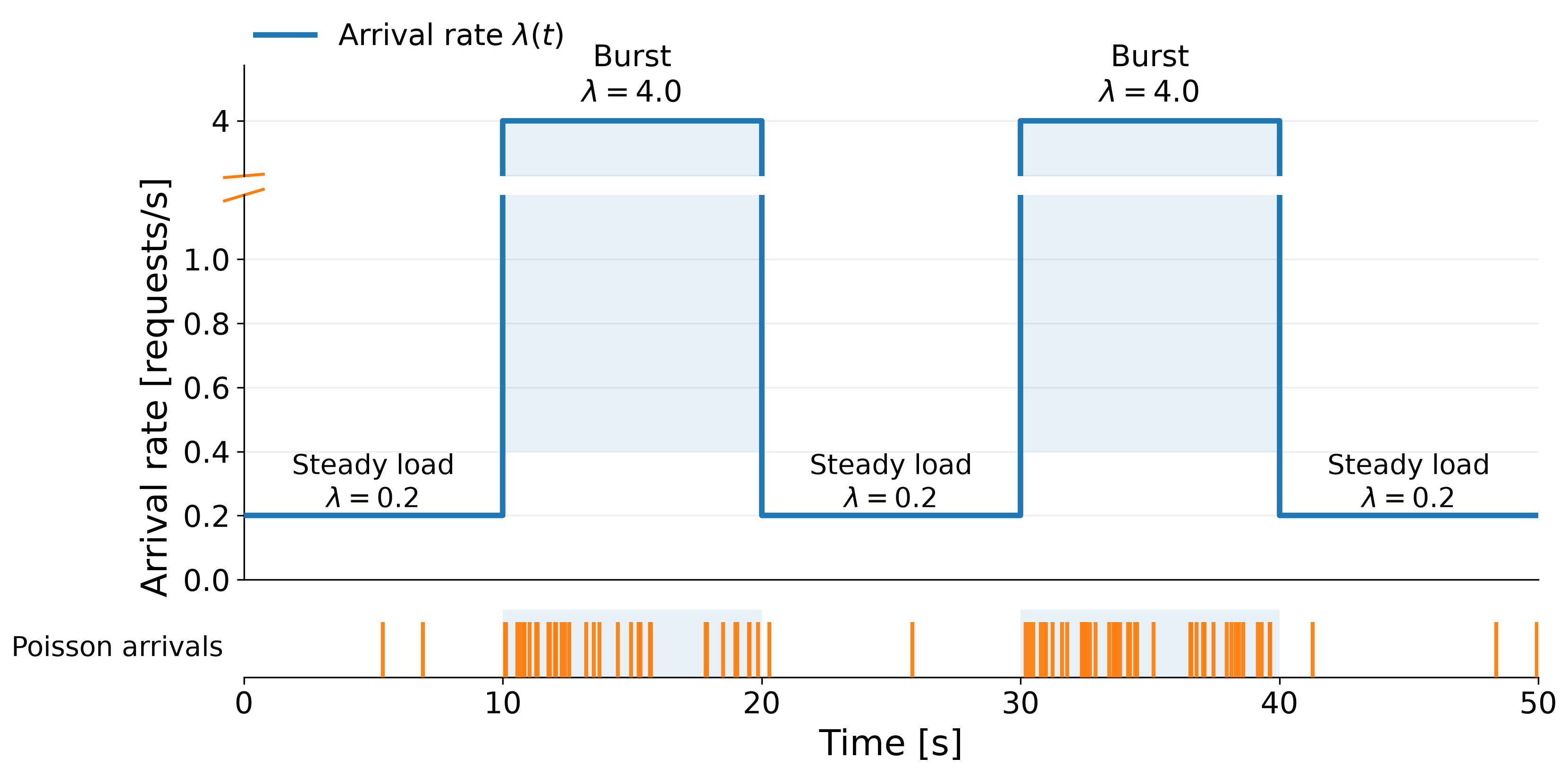}
    \caption{Bursty arrival workload used in the homogeneous P-PAS experiments. The arrival rate alternates between steady-load phases ($\lambda=0.2$ requests/s)
and higher-rate burst phases. Requests are sampled from a Poisson process
within each phase. The example shows a burst rate of
$\lambda=4.0$ requests/s; the y-axis is broken for readability.}
    \label{fig:workload}
\end{figure}

To evaluate scheduler behavior under changing load, we use 50\,s traces that
alternate between steady-load and burst phases. Request arrivals within each
phase follow a Poisson process with the corresponding arrival rate.
As illustrated in Figure~\ref{fig:workload}, each trace consists of 10\,s
phases, with a steady arrival rate of 0.2 requests/s and alternating burst
phases whose arrival rate is varied to control load. All homogeneous serving experiments in Section~\ref{sec:evaluation}
use this same temporal structure.

\subsection{Models and Hardware}

Our primary experiments use Qwen2.5-3B on an NVIDIA RTX 5090 (32\,GB) with 25k-token
prompts and 32-token outputs. We additionally evaluate Qwen2.5-0.5B and
SmolLM3-3B, vary prompt lengths from 20k to 30k tokens and output lengths
from 16 to 64 tokens, and perform experiments on NVIDIA A100 SXM (80\,GB) and H100 SXM
(80\,GB) GPUs. The upper prompt length of 30k was chosen close to the 32,768-token
context limit of Qwen2.5-3B, while the 20k--30k range provides multiple
operating points within the long-context regime. These variations test the
robustness of MBT sensitivity and P-PAS across model architectures and scales,
workload shapes, and consumer- and datacenter-class GPUs.

\subsection{Baselines and P-PAS Configuration}

We compare P-PAS against fixed MBT configurations of 2,048 and 16,384 tokens,
representing the small- and large-budget regimes identified in
Section~\ref{sec:mbt_sensitivity}. For P-PAS, we use
$N_{\mathrm{th}}=2$, $B_{\max}=16$k, and $B_{\mathrm{cap}}=2$k throughout
the homogeneous evaluation. These endpoints are selected from the primary static
sweep and then held fixed across the robustness experiments to test whether
the same policy transfers across models, workload shapes, and GPUs without
retuning.

\subsection{Metrics}

Our primary metric is end-to-end request latency, measured from request
arrival to completion of the generated response. We additionally report
time to first token (TTFT) and time per output token (TPOT) to characterize
prefill and decode behavior, respectively. Makespan measures the elapsed time
from the first request arrival until completion of the final request.

For request-level metrics, we report both the mean and the 95th percentile
(P95), where applicable. Unless stated otherwise, results for the homogeneous P-PAS experiments are averaged over five random seeds. 
The Poisson traces intentionally use the same seeds for all compared schedulers so
that each scheduler sees the same realized arrival sequence.

\section{Evaluation}
\label{sec:evaluation}

\subsection{P-PAS under Changing Load}

We first evaluate P-PAS under changing load using our primary configuration:
Qwen2.5-3B on the RTX 5090 with 25k-token prompts and 32-token outputs. We
vary the burst arrival rate from 0.4 to 1.4 requests/s while keeping the
steady arrival rate fixed at 0.2 requests/s.

\begin{figure}[t]
    \centering
    \includegraphics[width=\linewidth]{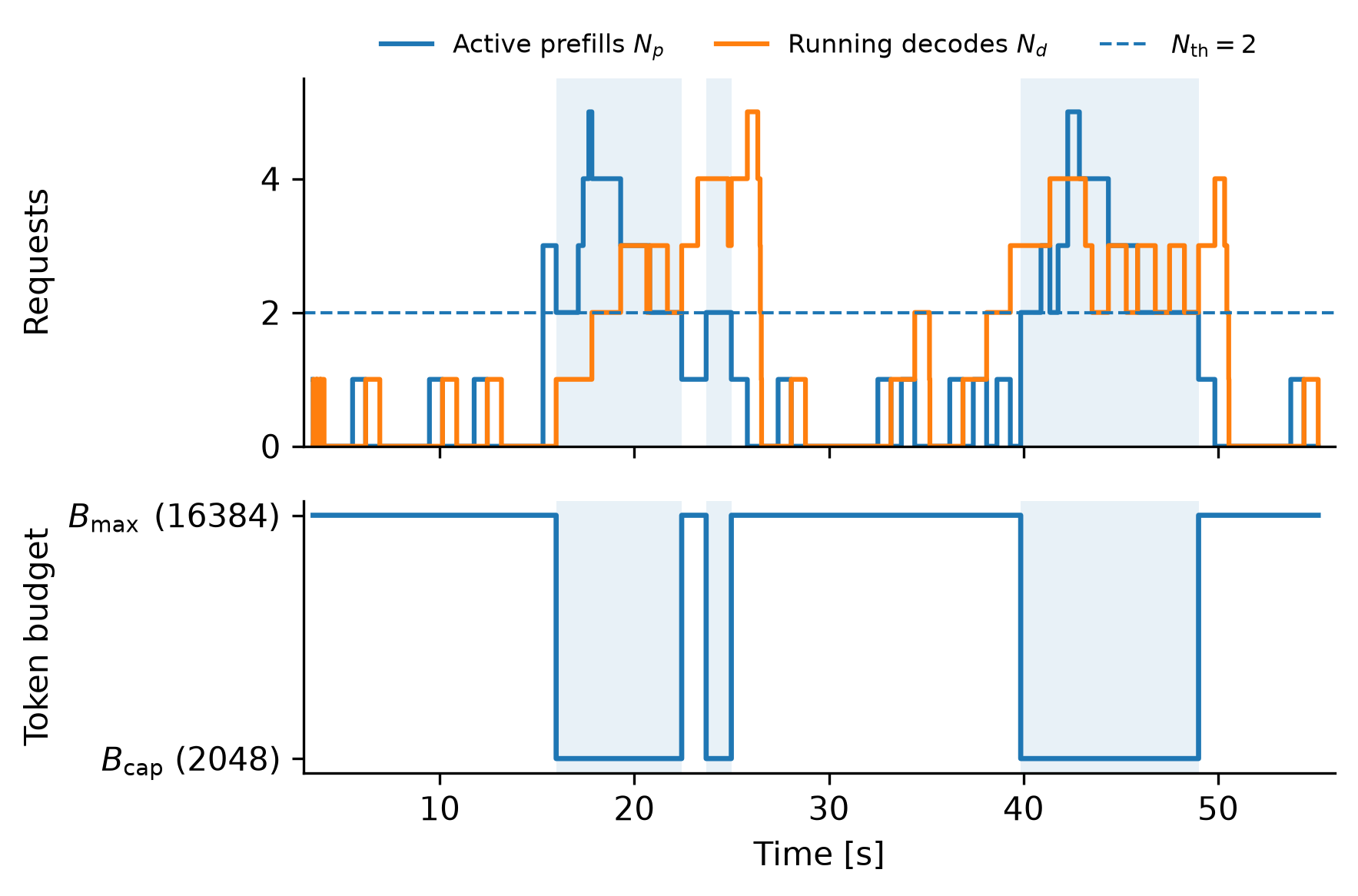}
    \caption{Representative P-PAS scheduler timeline from one dynamic
    Qwen2.5-3B run. The upper panel shows active prefills $N_p$ and running
    decodes $N_d$; the dashed line marks $N_{\mathrm{th}}=2$, and shaded
    intervals indicate high-pressure states. The lower panel shows the
    selected global token budget, which switches from
    $B_{\max}=16{,}384$ to $B_{\mathrm{cap}}=2{,}048$ under high pressure
    and returns to $B_{\max}$ when pressure subsides.}
    \label{fig:ppas_timeline}
\end{figure}

To make the within-run adaptation explicit, Fig.~\ref{fig:ppas_timeline}
shows one representative dynamic run at a burst arrival rate of 1.0 requests/s.
As the number of active prefills and running decodes changes, the scheduler
repeatedly enters and leaves high-pressure states. P-PAS correspondingly
switches between $B_{\max}$ and $B_{\mathrm{cap}}$. Thus, adaptation occurs
online within the same run rather than selecting a single budget for the
entire workload.

\begin{figure}[t]
    \centering
    \includegraphics[width=1.0\linewidth]{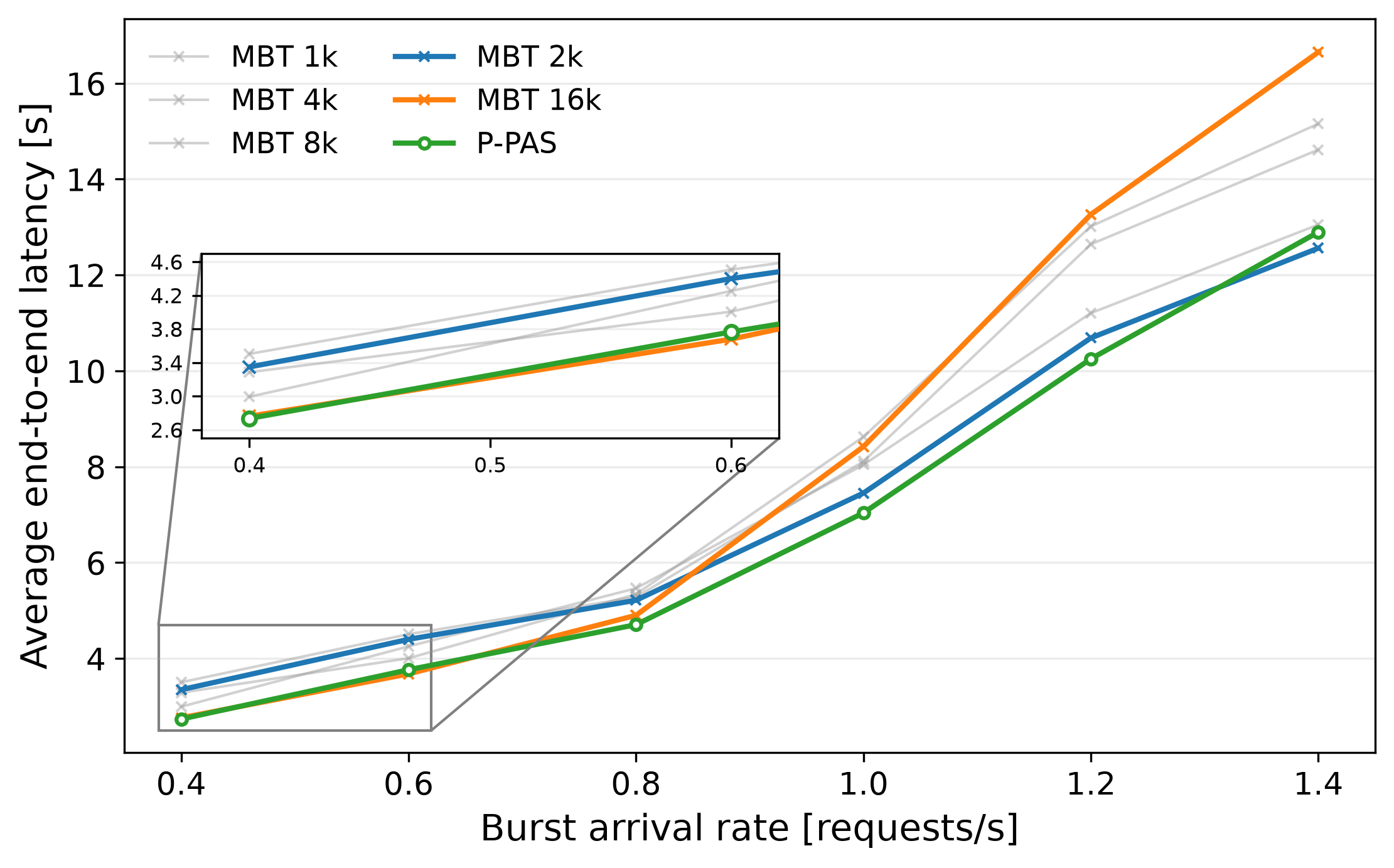}
    \caption{Average end-to-end latency across burst arrival rates for
    Qwen2.5-3B on an RTX 5090 with 25k-token prompts and 32-token outputs.
    Fixed MBT configurations favor different load regimes, whereas P-PAS
    adapts the scheduling token budget and maintains low latency as load
    changes. The inset highlights the low-load regime.}
    \label{fig:P-PAS_evaluation}
\end{figure}

Figure~\ref{fig:P-PAS_evaluation} compares P-PAS with fixed MBT configurations
across the full range of burst rates. Detailed per-burst results for all fixed
MBT configurations and P-PAS are provided in Appendix
Table~\ref{tab:qwen3b_full_results}. At low load, MBT 16k provides lower
end-to-end latency than MBT 2k, consistent with the MBT sensitivity identified
in Section~\ref{sec:mbt_sensitivity}. As load increases, the ordering reverses
and MBT 2k becomes preferable. P-PAS adapts across these regimes: it closely
matches MBT 16k at low load, outperforms both MBT 2k and MBT 16k across
intermediate burst rates, and remains close to MBT 2k at the highest load.
The additional static configurations (MBT 1k, 4k, and 8k) exhibit the same
limitation: none provides the lowest latency consistently across load regimes.

To summarize performance across all load regimes, we compare P-PAS with each
fixed-MBT baseline separately at every burst rate. For each metric, we first
compute the ratio between P-PAS and the baseline; a ratio below one indicates
that P-PAS performs better. We then aggregate these ratios across the $K$
burst rates using the geometric mean:

\begin{equation}
R_m =
\left(
\prod_{k=1}^{K}
\frac{m_{\mathrm{P\text{-}PAS},k}}
     {m_{\mathrm{baseline},k}}
\right)^{1/K},
\end{equation}

where $m$ denotes the evaluated metric. We express the resulting ratio as a
percentage improvement:

\begin{equation}
I_m = (1-R_m)\times100\%.
\end{equation}

Thus, positive values indicate an improvement over the fixed-MBT baseline. Table~\ref{tab:P-PAS_overall_5090}
summarizes the results across all six burst rates.

\begin{table}[t]
\centering
\caption{P-PAS improvement over fixed MBT schedulers for Qwen2.5-3B on the
RTX 5090. Values are geometric-mean improvements of normalized ratios across burst rates
0.4--1.4 requests/s. Positive values indicate lower values and better
performance.}
\label{tab:P-PAS_overall_5090}

\begin{tabular}{lcc}
\toprule
Metric & vs.\ MBT 2k & vs.\ MBT 16k \\
\midrule
Avg. latency & 8.5\% & 11.3\% \\
P95 latency  & 3.0\% & 10.0\% \\
Avg. TTFT    & 11.8\% & $-18.5\%$ \\
Avg. TPOT    & 3.4\% & 36.3\% \\
Makespan     & 1.0\% & $-3.9\%$ \\
\bottomrule
\end{tabular}
\end{table}

Across the full load range, P-PAS reduces average end-to-end latency by
8.5\% relative to MBT 2k and by 11.3\% relative to MBT 16k. The breakdown
also reflects the underlying scheduling trade-off: relative to MBT 16k,
P-PAS increases TTFT but substantially reduces TPOT, while relative to MBT 2k
it improves both. Makespan improves slightly relative to MBT 2k and increases
slightly relative to MBT 16k, consistent with the higher throughput generally
achieved by larger token budgets.

\subsection{Robustness across Models, Workloads, and GPUs}

We next test whether the observed behavior is specific to the primary
Qwen2.5-3B configuration. We vary model architecture and scale, prompt and
output length, and GPU platform while keeping the P-PAS policy and its
parameters unchanged.

To limit the number of experimental runs, the robustness experiments use
representative low- and high-pressure burst rates rather than the full
six-rate sweep of the primary evaluation. Aggregate improvements in this
section are therefore computed over the burst rates reported for each
experiment and can differ from the values obtained from the full sweep above.

\paragraph{Model architecture and scale.}
We evaluate Qwen2.5-0.5B, Qwen2.5-3B, and SmolLM3-3B on the RTX 5090.
Because Qwen2.5-0.5B processes requests substantially faster, higher burst
rates are required to expose comparable low- and high-pressure regimes.
Table~\ref{tab:ppas_models_5090} reports the geometric-mean improvement in
average end-to-end latency over the selected burst rates; the corresponding
absolute results are provided in Appendix
Table~\ref{tab:rtx5090_load_regimes}. P-PAS reduces average latency relative
to both fixed MBT baselines for all three models. The Qwen2.5-0.5B results
further show that the benefit is not restricted to models at the 3B scale.

\begin{table}[t]
\centering
\caption{Average-latency improvement of P-PAS across models on the RTX 5090.
Values are geometric-mean improvements across the indicated representative low- and
high-pressure burst rates. Positive values indicate lower latency.}
\label{tab:ppas_models_5090}

\setlength{\tabcolsep}{3.5pt}

\begin{tabular}{lccc}
\toprule
Model & Burst rates & vs.\ MBT 2k & vs.\ MBT 16k \\
\midrule
Qwen2.5-0.5B & 1.0, 4.0 & 7.1\%  & 14.2\% \\
Qwen2.5-3B   & 0.4, 1.0 & 12.2\% & 9.0\%  \\
SmolLM3-3B   & 0.4, 1.0 & 13.9\% & 9.0\%  \\
\bottomrule
\end{tabular}
\end{table}

\paragraph{Prompt and output length.}
We next vary the workload shape for Qwen2.5-3B. As shown in
Table~\ref{tab:robustness_prompt_output}, P-PAS reduces average latency relative to both
fixed MBT baselines across all evaluated prompt and output lengths.
The magnitude of the improvement varies with workload shape, reaching
23.2\% relative to MBT 16k for 16-token outputs. P95 latency follows the
same general trend, with one exception: for 30k-token prompts, P-PAS is
2.7\% worse than MBT 2k while remaining 7.1\% better than MBT 16k.
The results further show that the benefit of P-PAS is not restricted to the
25k-token prompt and 32-token output configuration used in our primary
experiments.

\begin{table}[t]
\centering
\caption{Robustness of P-PAS across prompt and output lengths for
Qwen2.5-3B. Values are geometric-mean latency improvements across burst
rates 0.8 and 1.2. Positive values indicate lower latency.}
\label{tab:robustness_prompt_output}

\setlength{\tabcolsep}{3.5pt}

\begin{tabular}{llrrrr}
\toprule
Experiment & Value &
\multicolumn{2}{c}{vs.\ MBT 2k} &
\multicolumn{2}{c}{vs.\ MBT 16k} \\
\cmidrule(lr){3-4}\cmidrule(lr){5-6}
&& Avg. & P95 & Avg. & P95 \\
\midrule

\multirow{3}{*}{Prompt}
& 20k & 13.2\% & 8.4\%  & 11.6\% & 11.5\% \\
& 25k & 7.0\%  & 1.4\%  & 13.9\% & 14.4\% \\
& 30k & 6.3\%  & $-2.7$\% & 15.2\% & 7.1\% \\

\midrule

\multirow{3}{*}{Output}
& 16 & 5.9\% & 1.3\% & 23.2\% & 20.1\% \\
& 32 & 7.0\% & 1.4\% & 13.9\% & 14.4\% \\
& 64 & 8.4\% & 5.4\% & 1.8\%  & 3.4\% \\

\bottomrule
\end{tabular}
\end{table}

\paragraph{GPU platform.}
We next evaluate P-PAS on an NVIDIA A100 SXM to determine whether its benefits
extend from the consumer-class RTX 5090 to datacenter hardware.
Table~\ref{tab:ppas_a100} shows that P-PAS reduces average end-to-end latency
relative to both fixed MBT baselines for both evaluated models. For
Qwen2.5-3B, P-PAS improves average latency by 7.8\% over MBT 2k and 6.6\%
over MBT 16k. For SmolLM3-3B, the corresponding improvements are 10.2\%
and 5.4\%. These results show that the benefits observed on the Blackwell-based RTX 5090 persist on the earlier-generation Ampere-based A100 datacenter GPU. Absolute per-burst measurements for both models on the A100 SXM are provided in
Appendix Table~\ref{tab:a100_detailed_results}.

\begin{table}[t]
\centering
\caption{P-PAS performance on the NVIDIA A100 SXM. Values are
geometric-mean improvements over the corresponding fixed-MBT baseline
across representative low- and high-pressure burst rates 0.4 and 1.0.
Positive values indicate lower values and better performance.}
\label{tab:ppas_a100}

\setlength{\tabcolsep}{3.5pt}

\begin{tabular}{llcc}
\toprule
Model & Metric & vs.\ MBT 2k & vs.\ MBT 16k \\
\midrule

\multirow{5}{*}{Qwen2.5-3B}
& Avg. latency & 7.8\% & 6.6\% \\
& P95 latency  & 7.0\% & 14.8\% \\
& Avg. TTFT    & 7.8\% & $-5.6\%$ \\
& Avg. TPOT    & 8.2\% & 13.6\% \\
& Makespan     & 0.4\% & $-1.6\%$ \\

\midrule

\multirow{5}{*}{SmolLM3-3B}
& Avg. latency & 10.2\% & 5.4\% \\
& P95 latency  & 10.6\% & 7.3\% \\
& Avg. TTFT    & 10.6\% & $-11.7\%$ \\
& Avg. TPOT    & 9.7\% & 20.8\% \\
& Makespan     & 0.7\% & $-1.5\%$ \\

\bottomrule
\end{tabular}
\end{table}

Overall, these experiments show that the benefits of P-PAS persist across
model architectures and scales, workload shapes, and both consumer- and
datacenter-class GPUs. The magnitude of the improvement varies across
configurations, consistent with MBT sensitivity depending on the interaction
between model, workload, and hardware. We additionally verify that the underlying load-dependent MBT crossover occurs
on an H100 SXM; detailed results are provided in Appendix
Table~\ref{tab:h100}. Additional homogeneous large-model results are provided
in Appendix~\ref{app:large_model_homogeneous}.

\section{Understanding MBT Sensitivity}
\label{sec:understanding_mbt}

We next investigate the mechanisms underlying MBT sensitivity using GPU
profiling with NVIDIA Nsight Systems and Nsight Compute. We first analyze
why large token budgets can improve prefill execution under low scheduling
pressure and why this advantage varies across serving configurations
(Section~\ref{sec:prefill_efficiency}). We then examine why smaller token
budgets become preferable as scheduling pressure increases
(Section~\ref{sec:decode_interference}).

\subsection{Large-Chunk Prefill Efficiency and Its Limits}
\label{sec:prefill_efficiency}

Under low scheduling pressure, there is little concurrent decode work for
large prefill chunks to interfere with. In this regime, the latency difference
between MBT configurations is therefore strongly influenced by the efficiency
with which the prefill itself is executed.

\begin{figure}[t]
    \centering
    \begin{subfigure}{0.47\linewidth}
        \centering
        \includegraphics[width=\linewidth]{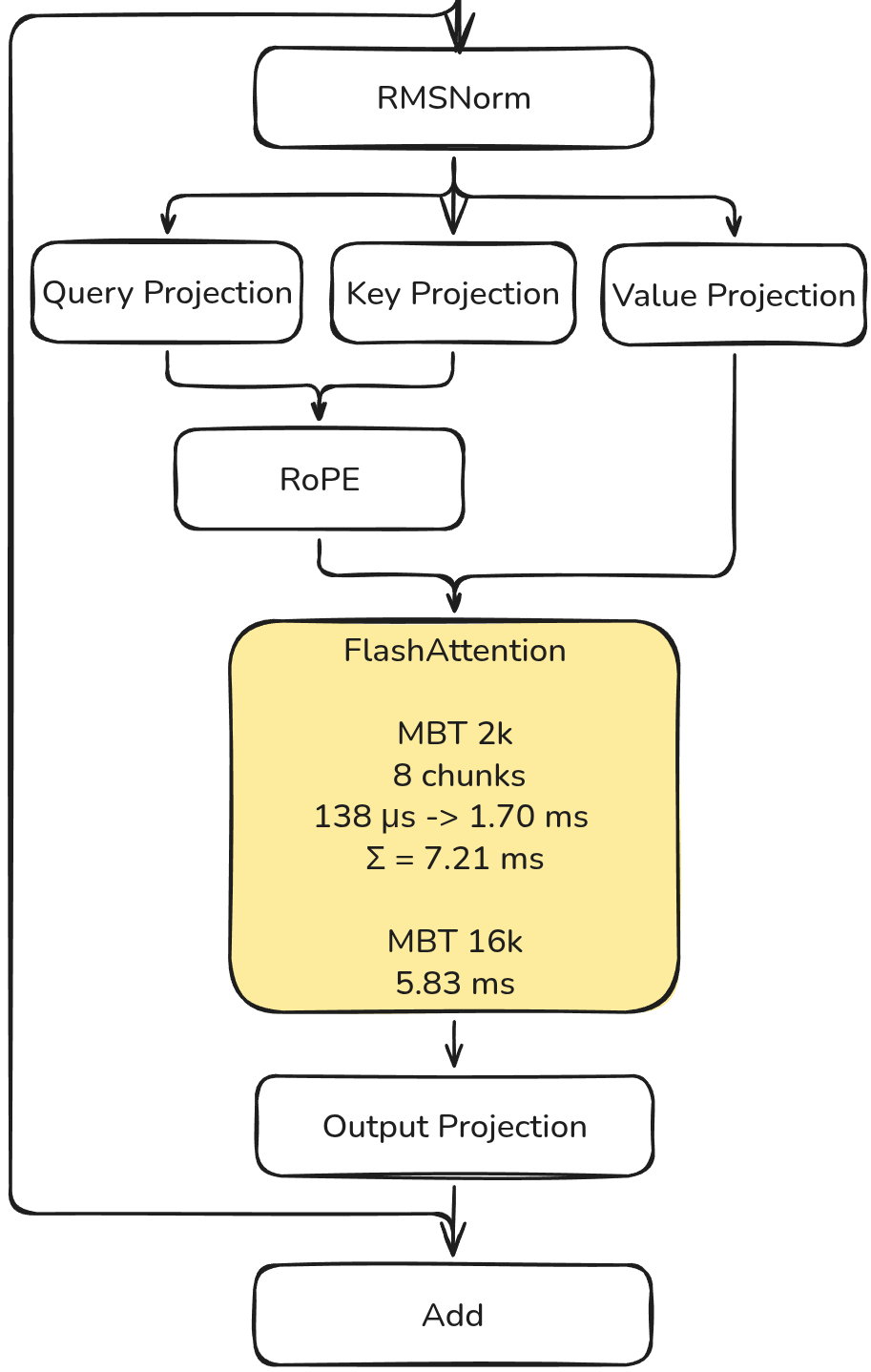}
        \caption{Attention}
        \label{fig:kernel_attention}
    \end{subfigure}
    \hspace{0.03\linewidth}
    \begin{subfigure}{0.33\linewidth}
        \centering
        \includegraphics[width=\linewidth]{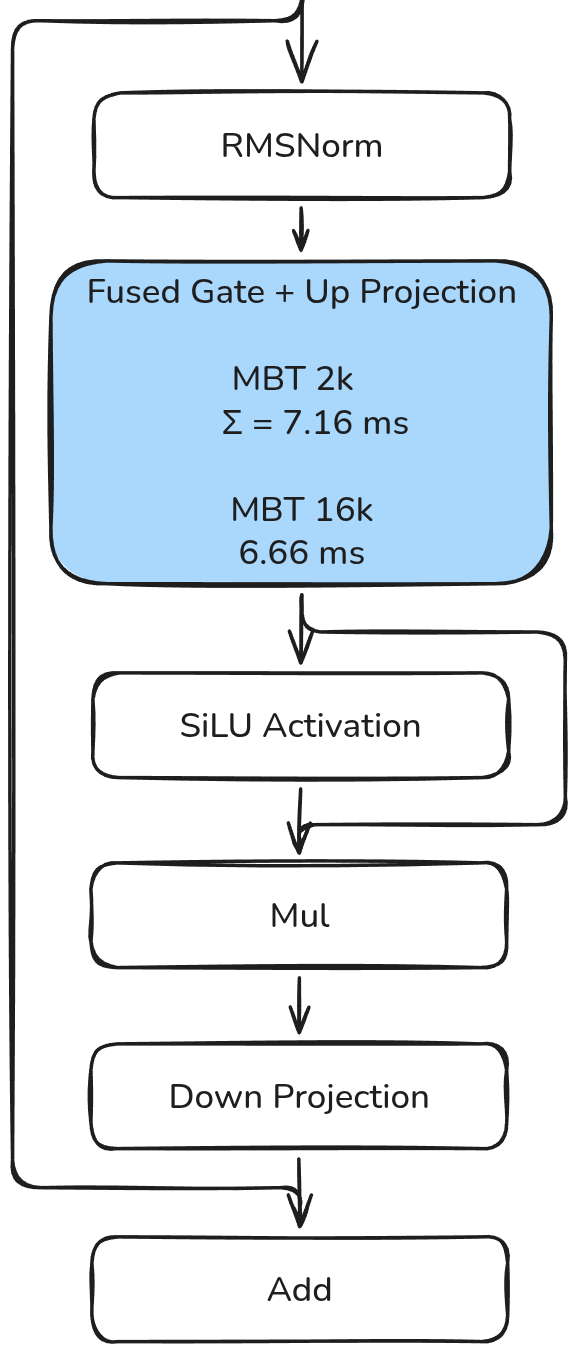}
        \caption{MLP}
        \label{fig:kernel_mlp}
    \end{subfigure}

    \caption{Kernel-level comparison of small- and large-chunk prefill
    execution for one Qwen2.5-3B decoder block. Processing the same 16,384-token prompt
    through eight 2k chunks incurs a higher cumulative execution time
    for dominant attention and MLP operations than processing it as a
    single 16k chunk.}
    \label{fig:qwen_kernel_breakdown}
\end{figure}

Figure~\ref{fig:qwen_kernel_breakdown} illustrates this effect for one
Qwen2.5-3B decoder block using kernel execution times measured with Nsight
Systems. We compare processing the same number of prompt tokens using
multiple 2k chunks with processing them as a single 16k chunk. The figure
highlights FlashAttention and the fused gate-up projection, two
computationally expensive operations during prefill.

For FlashAttention, the eight 2k chunks require 7.21\,ms in total, compared
with 5.83\,ms for a single 16k chunk. This cumulative cost includes the
progressively increasing execution time of the individual 2k chunks as the
KV sequence grows. The fused gate-up projection shows a smaller but similar
effect: its cumulative execution time decreases from 7.16\,ms for the 2k
execution to 6.66\,ms for 16k. Thus, larger prefill chunks can improve the
execution efficiency of dominant prefill operations rather than merely
reducing the number of scheduler iterations.

We next examine whether this kernel-level advantage translates to the complete
prefill and whether it persists across model sizes.
Table~\ref{tab:prefill_scaling} reports total prefill execution time for
three Qwen2.5 models.

\begin{table}[t]
\centering
\caption{Total prefill execution time for Qwen2.5 models on the RTX 5090,
measured with Nsight Systems. The final column reports the reduction when
using MBT 16k instead of MBT 2k.}
\label{tab:prefill_scaling}

\setlength{\tabcolsep}{3.5pt}

\begin{tabular}{lrrr}
\toprule
Model & MBT 2k [ms] & MBT 16k [ms] & Reduction \\
\midrule
Qwen2.5-0.5B & 148.0 & 132.4 & 10.5\% \\
Qwen2.5-3B   & 755.8 & 668.6  & 11.6\% \\
Qwen2.5-7B   & 1373  & 1365  & 0.6\% \\
\bottomrule
\end{tabular}
\end{table}

For Qwen2.5-0.5B and Qwen2.5-3B, increasing the MBT from 2k to 16k reduces
total prefill time by 10.5\% and 11.6\%, respectively. This provides a
meaningful execution-efficiency advantage for the large MBT under low
scheduling pressure. For Qwen2.5-7B, however, the reduction is only 0.6\%.
A broader MBT sweep still shows the expected pressure-dependent crossover:
large MBTs are preferred at low pressure, whereas MBTs around 1k--2k perform
best under high pressure. However, the low-pressure latency surface is shallow,
with MBT 16k improving average latency by only about 1\% over the 1k--2k
region. P-PAS therefore has comparatively little additional latency benefit
to recover for this configuration.

To investigate why the large-chunk advantage diminishes for Qwen2.5-7B, we
profile the dominant fused gate-up projection with Nsight Compute.
Table~\ref{tab:ncu_scaling} reports SM throughput, L2 hit rate, and DRAM
throughput for MBT 2k and MBT 16k. Here, SM throughput reflects compute
utilization relative to the GPU's sustained peak, L2 hit rate indicates the
fraction of L2 cache accesses served by the cache, and DRAM throughput
reflects memory traffic relative to sustained DRAM bandwidth.

\begin{table}[t]
\centering
\caption{Nsight Compute measurements for the dominant gate-up projection
during prefill on the RTX 5090. Throughput values are reported as percentages
of the corresponding sustained peak.}
\label{tab:ncu_scaling}

\setlength{\tabcolsep}{3.5pt}

\begin{tabular}{lrrrr}
\toprule
Model & MBT & SM [\%] & L2 Hit [\%] & DRAM [\%] \\
\midrule
\multirow{2}{*}{Qwen2.5-0.5B}
& 2k  & 40.64 & 86.95 & 12.37 \\
& 16k & 45.76 & 90.03 & 11.91 \\
\midrule
\multirow{2}{*}{Qwen2.5-3B}
& 2k  & 43.64 & 91.73 & 8.77 \\
& 16k & 48.06 & 90.53 & 11.38 \\
\midrule
\multirow{2}{*}{Qwen2.5-7B}
& 2k  & 47.92 & 96.04 & 7.69 \\
& 16k & 48.40 & 88.95 & 22.14 \\
\bottomrule
\end{tabular}
\end{table}

For the two smaller models, increasing the prefill chunk size improves SM
throughput. Qwen2.5-0.5B increases from 40.64\% to 45.76\%, while its L2
hit rate increases and DRAM throughput remains approximately unchanged.
Qwen2.5-3B similarly increases SM throughput from 43.64\% to 48.06\%,
with only modest changes in L2 hit rate and DRAM throughput.

Qwen2.5-7B exhibits a different behavior. Increasing the MBT from 2k to 16k
barely changes SM throughput, from 47.92\% to 48.40\%, while the L2 hit rate
decreases from 96.04\% to 88.95\% and DRAM throughput increases from 7.69\%
to 22.14\%. The larger prefill operation therefore places substantially more
pressure on the memory hierarchy without translating this into higher SM
throughput. This behavior coincides with the near disappearance of the
whole-prefill advantage of MBT 16k in Table~\ref{tab:prefill_scaling}.

These results show that large-chunk prefill efficiency is one contributor to
the low-pressure MBT advantage and depends on the execution characteristics
of the particular model--hardware configuration. For Qwen2.5-0.5B and
Qwen2.5-3B on the RTX 5090, moving to larger chunks provides a clear
prefill-efficiency gain. Qwen2.5-7B still exhibits a pressure-dependent MBT
optimum, but the low-pressure latency advantage of a large MBT is much smaller,
reducing the headroom available to adaptive scheduling.

\subsection{Decode Interference under Scheduling Pressure}
\label{sec:decode_interference}

The preceding analysis explains why a large MBT can be advantageous under
low scheduling pressure. It does not, however, explain why the ordering
reverses as pressure increases.

Under higher scheduling pressure, multiple requests can require prefill while
other requests are already decoding. vLLM can schedule prefill and decode
tokens together within the same iteration. With a large MBT, however, the
decode work can be accompanied by a large amount of prefill work, increasing
the execution time of the iteration. Although active decode requests make
progress in that iteration, their next decode step cannot be scheduled until
the iteration completes.

\begin{figure}[t]
    \centering
    \includegraphics[width=\linewidth]{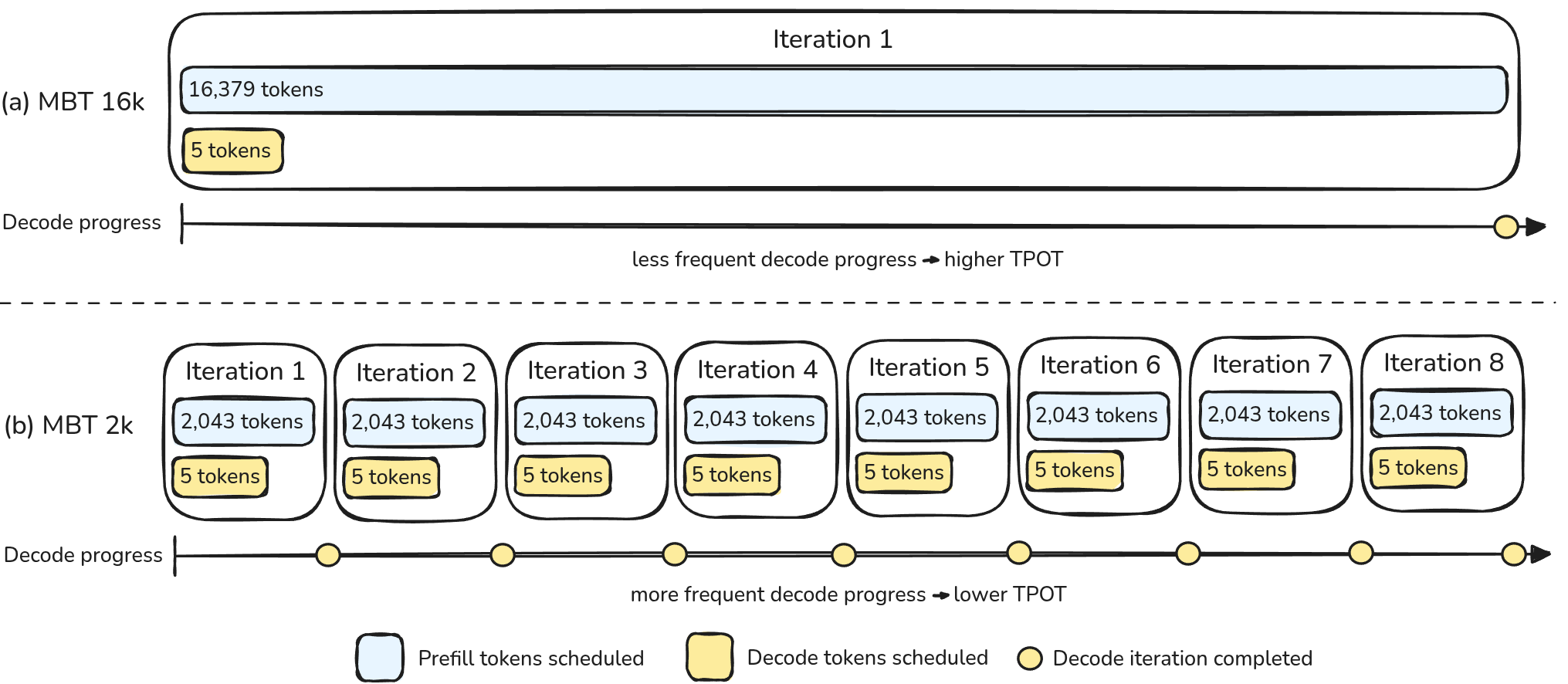}
\caption{Prefill--decode interference under scheduling pressure.
Prefill and decode tokens share the MBT within each iteration.
(a) MBT 16k processes the prefill in one large iteration, resulting in
less frequent decode progress. (b) MBT 2k distributes approximately the
same prefill work across eight iterations, enabling more frequent decode
progress and lower TPOT.}
    \label{fig:decode_interference}
\end{figure}

Figure~\ref{fig:decode_interference} illustrates this interaction
conceptually. With MBT 16k, a large amount of prefill work can be scheduled
alongside the active decode requests, resulting in fewer but longer
iterations. With MBT 2k, approximately the same prefill work is distributed
across multiple smaller iterations. Decode requests therefore make progress
more frequently.

This more frequent decode progress improves TPOT, but requires long prefills
to be distributed across more iterations and can therefore increase TTFT.
Under sufficient scheduling pressure, the TPOT improvement can outweigh this
TTFT disadvantage, making MBT 2k preferable for end-to-end latency.

MBT sensitivity therefore results from two competing effects. A larger MBT
can improve prefill execution efficiency, while a smaller MBT limits
prefill--decode interference under scheduling pressure. A pronounced
load-dependent crossover arises when both effects are sufficiently strong.
P-PAS exploits this crossover by retaining the larger scheduling budget under
low pressure and constraining it as prefill pressure increases. Which endpoint
pair best exposes this crossover is deployment-dependent, motivating the
calibration procedure in Section~\ref{sec:calibration}.

\section{Extending P-PAS to Heterogeneous Requests: P-PAS+}
\label{sec:ppas_plus}

The main P-PAS experiments intentionally use homogeneous request shapes to
isolate the pressure-dependent MBT crossover. Real serving workloads, however, can mix short and very long prompts
as well as different output lengths~\citep{xiang2026servegen,ruan2026libra}. Under
such heterogeneity, request count alone can become an incomplete proxy for
prefill pressure: one remaining 128k-token prefill can represent substantially
more work than several shorter prefills.

P-PAS+ extends the P-PAS pressure detector while leaving the budget-switching
mechanism unchanged. It retains the count-based signal in
Eq.~\ref{eq:count_pressure} and adds a residual-work condition for the case in
which one large running prefill overlaps with multiple active decodes. Let
$R_p$ denote the number of remaining tokens of the running prefill and
$T_{\mathrm{large}}$ a deployment-specific threshold. We define
\begin{equation}
P_{\mathrm{large}}
=
(N_{\mathrm{prefill}}^{\mathrm{running}}=1)
\land
(N_d\ge N_{d,\mathrm{large}})
\land
(R_p\ge T_{\mathrm{large}}),
\end{equation}
and use
\begin{equation}
P_{\mathrm{P\text{-}PAS+}}
=
P_{\mathrm{count}}\lor P_{\mathrm{large}}.
\label{eq:ppas_plus_pressure}
\end{equation}

P-PAS+ then uses the same budget endpoints as P-PAS: $B_{\mathrm{cap}}$
under pressure and $B_{\max}$ otherwise. We use
$N_{d,\mathrm{large}}=2$ in the experiments below.

The distinction between P-PAS and P-PAS+ is intentionally narrow. P-PAS uses
the count-based pressure signal in Eq.~\ref{eq:count_pressure} and is sufficient
for the homogeneous workloads in the main study. P-PAS+ adds one signal that becomes informative when request
sizes vary substantially. In a deployment, P-PAS+ can be used generally; when
the large-prefill condition never fires, it reduces to P-PAS
behavior. We retain the separate name in the paper to make the heterogeneous
extension and its ablation explicit.

\subsection{Heterogeneous Evaluation Setup}

We evaluate P-PAS+ on two substantially larger recent models,
Qwen3.8-27B-NVFP4 and NVIDIA Nemotron-3.5-Lightning-30B-A3B-NVFP4.
We selected these models because they are representative recent models
for the long-context, agentic workloads targeted by P-PAS+, while also
providing substantially larger model configurations than the Qwen2.5
models used in the main P-PAS evaluation.

Unlike the earlier homogeneous experiments, each run contains a mixture of
prompt and output lengths. For Qwen, we use prompt lengths
$\{8\mathrm{k},16\mathrm{k},32\mathrm{k},64\mathrm{k}\}$, while Nemotron
uses $\{16\mathrm{k},32\mathrm{k},64\mathrm{k},128\mathrm{k}\}$. Both
models use output lengths $\{16,32,64,128,256\}$ tokens. The different prompt
ranges reflect the maximum practical context lengths on the RTX 5090 used in
our experiments: Nemotron supports 128k-token prompts, whereas Qwen runs out
of GPU memory at that length. We therefore use 8k--64k for Qwen to retain four
prompt-length levels across both models.

A small static MBT sweep is used to characterize each deployment. For Qwen we
compare MBT $\{512,768,1024,2048,4096,8192\}$ and use
$B_{\max}=2048$ and $B_{\mathrm{cap}}=768$. For Nemotron we compare MBT
$\{1024,1280,2048,4096,8192,16384\}$ and use
$B_{\max}=16384$ and $B_{\mathrm{cap}}=1280$. The best tested static
configuration for mean end-to-end latency is MBT 768 for Qwen and MBT 1280
for Nemotron.

For the residual-prefill condition, we set $T_{\mathrm{large}}$ to half of
the largest prompt length used for each model, giving 32k tokens for Qwen and
64k for Nemotron. This provides a simple scale-relative criterion for
identifying a single unusually large remaining prefill. Setting the threshold
near the maximum prompt length would make the additional condition rarely
useful, while lower thresholds increasingly overlap with ordinary single-
prefill execution. We do not tune this threshold further in the present
evaluation; its sensitivity is left for future study. 
The resulting
P-PAS+ configurations are summarized in Table~\ref{tab:ppas_plus_config}.
\ref{app:heterogeneous_schedules} gives the complete schedules and aggregation. The evaluation
required 15 GPU-hours for Qwen and 18 for Nemotron (33 total).

\begin{table}[t]
\centering
\caption{P-PAS+ configurations for the heterogeneous evaluation.}
\label{tab:ppas_plus_config}

\setlength{\tabcolsep}{3.5pt}

\begin{tabular}{lrrr}
\toprule
Model & $B_{\max}$ & $B_{\mathrm{cap}}$ & $T_{\mathrm{large}}$ \\
\midrule
Qwen3.8-27B & 2048 & 768 & 32768 \\
Nemotron-30B & 16384 & 1280 & 65536 \\
\bottomrule
\end{tabular}
\end{table}

\subsection{Heterogeneous Results}

P-PAS+ improves over the best tested static configuration on both models.
For Qwen3.8-27B, it reduces mean end-to-end latency by 3.02\% relative to
MBT 768, with improvements of 3.01\% in P95 latency and 1.57\% in makespan.
For Nemotron-30B, it improves mean latency by 3.1\% and makespan by 1.4\%
relative to MBT 1280. Table~\ref{tab:heterogeneous_results} summarizes all
static comparisons, including a 17.0\% mean-latency improvement over MBT 2048
on Nemotron.

\begin{figure}[t]
    \centering
    \begin{subfigure}[t]{0.47\linewidth}
        \centering
        \maybeincludegraphics[width=\linewidth]
        {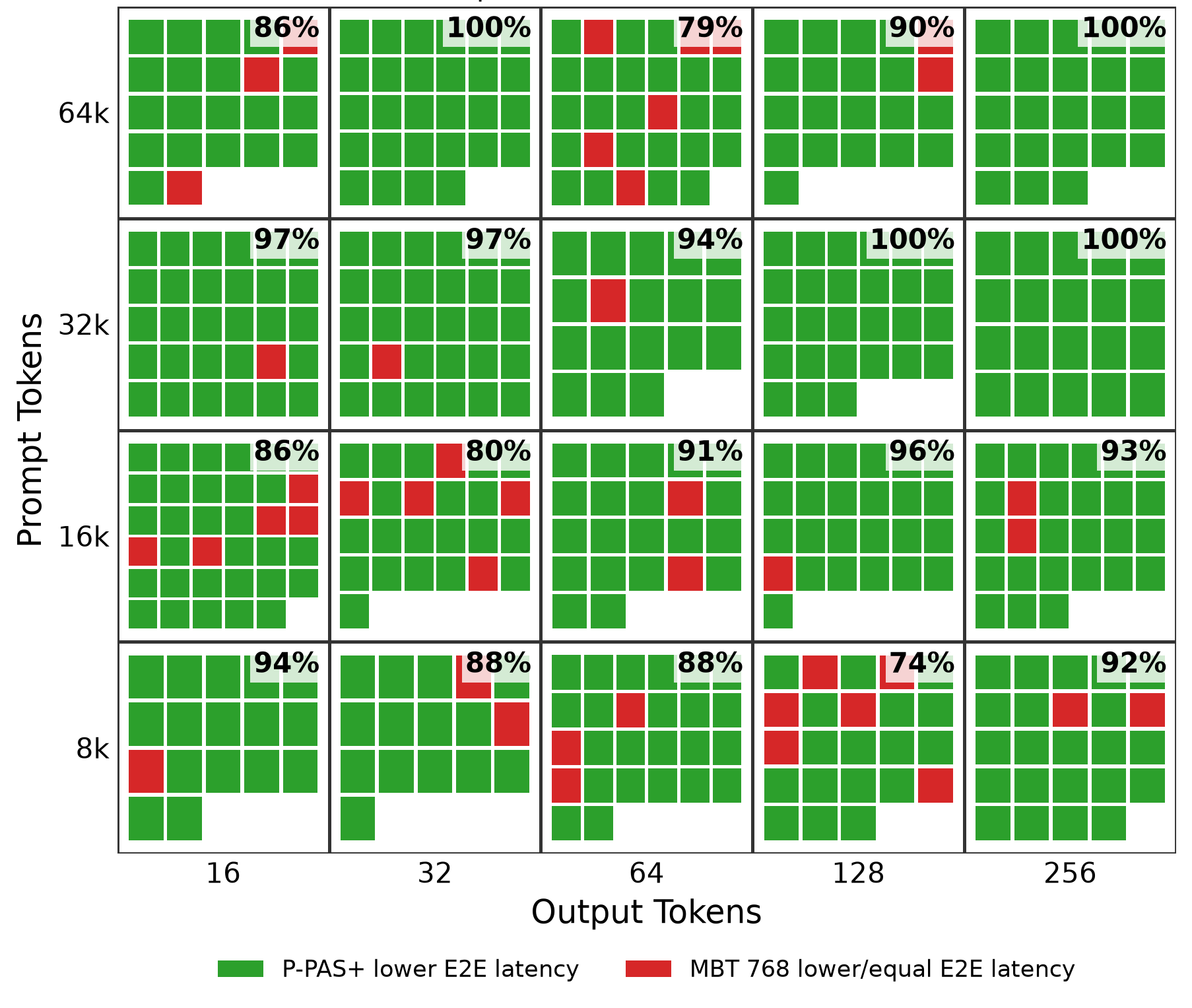}
        \caption{Qwen3.8-27B: P-PAS+ vs.\ MBT 768.}
        \label{fig:qwen_request_wall}
    \end{subfigure}
    \hfill
    \begin{subfigure}[t]{0.47\linewidth}
        \centering
        \maybeincludegraphics[width=\linewidth]
        {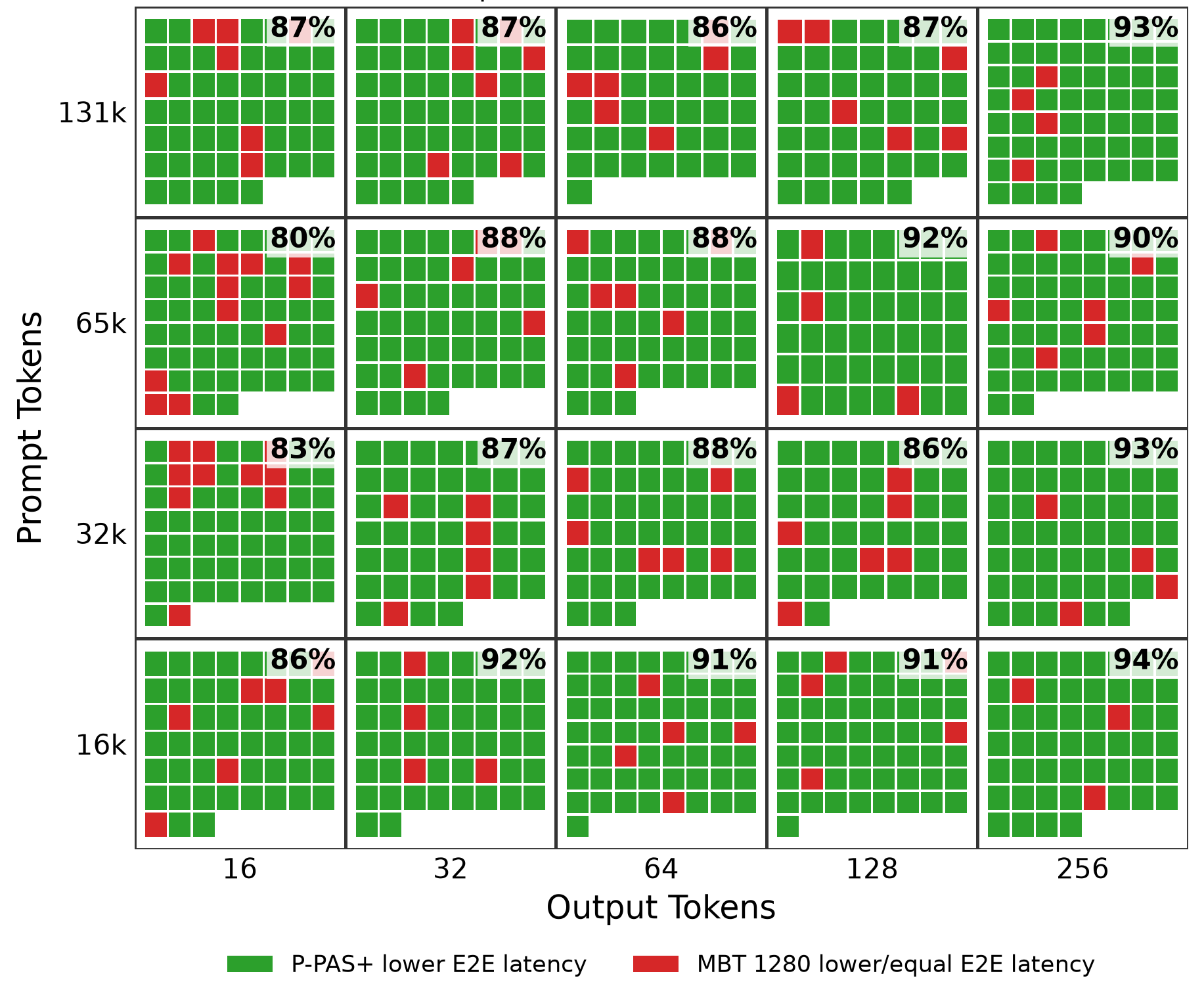}
        \caption{Nemotron-30B: P-PAS+ vs.\ MBT 1280.}
        \label{fig:nemotron_request_wall}
    \end{subfigure}
    \caption{Matched-request comparison of P-PAS+ against the best tested
    static configuration. Each point represents the same logical request
    under both schedulers. P-PAS+ is faster for 91.2\% of 488 matched Qwen
    requests and 88.5\% of 1,045 matched Nemotron requests.}
    \label{fig:heterogeneous_request_wall}
\end{figure}

Relative to P-PAS, P-PAS+ further reduces mean latency by 0.56\% on Qwen
and 1.4\% on Nemotron. Figure~\ref{fig:heterogeneous_request_wall} shows
that the gains are broadly distributed: 445 of 488 matched Qwen requests
(91.2\%) and 88.5\% of 1,045 matched Nemotron requests complete faster
than with the respective best static configuration.

\begin{table}[t]
\centering
\caption{Mean end-to-end latency improvement of P-PAS+ over static budgets
and P-PAS. Positive values indicate lower latency.}
\label{tab:heterogeneous_results}
\small
\setlength{\tabcolsep}{4pt}
\begin{tabular}{llr}
\toprule
Model & Comparison & Improvement \\
\midrule
\multirow{7}{*}{Qwen3.8-27B}
& MBT 512 & 23.55\% \\
& MBT 768 (best static) & 3.02\% \\
& MBT 1024 & 9.68\% \\
& MBT 2048 & 3.32\% \\
& MBT 4096 & 7.14\% \\
& MBT 8192 & 4.45\% \\
& P-PAS & 0.56\% \\
\midrule
\multirow{7}{*}{Nemotron-30B}
& MBT 1024 & 14.9\% \\
& MBT 1280 (best static) & 3.1\% \\
& MBT 2048 & 17.0\% \\
& MBT 4096 & 16.4\% \\
& MBT 8192 & 18.4\% \\
& MBT 16384 & 21.2\% \\
& P-PAS & 1.4\% \\
\bottomrule
\end{tabular}
\end{table}

The smaller gain over P-PAS is expected because P-PAS+ changes only the
pressure detector: P-PAS already handles multiple active prefills, while
the additional condition covers large single-prefill overlap with active
decoding. The same pressure-adaptive principle therefore remains effective
under heterogeneous request shapes and on recent 27B/30B models.

\subsection{Limitations and Scope}
\label{sec:limitations}

P-PAS is most useful when different pressure regimes favor meaningfully
different token budgets. Qwen2.5-7B shows that such a crossover can exist
while the latency surface remains shallow, leaving little headroom over a
well-chosen static MBT. We currently identify this headroom and select the
budget endpoints empirically rather than predicting them from model and
hardware characteristics.

Our evaluation primarily targets long-context, short-output serving, where
prefill--decode interference is pronounced. Two budget endpoints capture
nearly all of the available benefit in the evaluated short-output regime,
while longer decode-heavy workloads can favor intermediate budgets. Finally,
experiments use single-GPU model instances and controlled arrival processes;
tensor-parallel or multi-replica serving, production traces, and interactions
with mechanisms such as speculative decoding remain future work.

\section{Related Work}

LLM serving systems dynamically batch requests at different stages of
autoregressive inference to efficiently serve concurrent workloads. Orca~\cite{yu2022orca}
introduced iteration-level scheduling, allowing the composition of a batch to
change between decoding iterations. vLLM~\cite{kwon2023efficient} combines
continuous batching with PagedAttention to efficiently manage the dynamically
growing KV cache. 

Several systems specifically address the interaction between prefill and
decode. Sarathi~\cite{agrawal2023sarathi} introduced chunked prefill and
decode-maximal batching, which combine prefill chunks with active decode
requests to improve serving efficiency. DeepSpeed-FastGen~\cite{holmes2024deepspeed}
pursues a related approach with Dynamic SplitFuse, which splits and combines
prompt and generation work into more uniformly sized forward passes.
Sarathi-Serve~\cite{agrawal2024taming} develops chunked-prefill scheduling
into a stall-free serving scheduler that prioritizes ongoing decodes and
fills the remaining token budget with prefill work, limiting prefill-induced
generation stalls.

Chunked prefill has since been integrated into vLLM, where
\texttt{max\_num\_batched\_tokens} (MBT) limits the total token budget
available to prefill and decode work within an iteration. Unlike
Sarathi-Serve, which uses a pre-configured chunk size, P-PAS adapts the
scheduling token budget online based on the current scheduler state.

More recent systems explore alternative ways of controlling prefill execution.
\citet{liu2026fairness} regulate prefill concurrency and scheduling
based on fairness and latency objectives. FlowPrefill~\cite{hsieh2026flowprefill}
enables preemption at operator boundaries, allowing large prefill chunks to be
interrupted to protect responsiveness, while Layered Prefill~\cite{lee2026tokens} partitions prefill along
transformer layers to reduce the weight-loading overhead of token-level
chunking for MoE models. These
approaches address related prefill scheduling and execution trade-offs but
differ from our focus on the load-dependent MBT crossover and online
adaptation of the scheduling token budget.
Alternatively, Splitwise~\cite{patel2024splitwise} and
DistServe~\cite{zhong2024distserve} avoid colocated prefill--decode
interference by executing the two phases on separate hardware resources.

Recent systems also target request-shape heterogeneity. LAPS~\cite{she2026laps}
uses prompt length to adapt prefill scheduling, while Libra~\cite{ruan2026libra}
dynamically partitions requests under variable prompt and response
lengths. P-PAS+ instead augments the scheduling-pressure signal with
remaining prefill work while retaining the same budget-switching mechanism.

Beyond prefill--decode scheduling specifically, recent work has shown that
inference-engine configuration effects can be strongly model- and
workload-dependent. \citet{zine2026attention} systematically
evaluate attention kernels, prefix caching, and chunked prefill in vLLM,
finding that no evaluated configuration is universally optimal across models
and workloads. In their setting, the observed effects are primarily driven
by attention type and prefix caching, while enabling chunked prefill has
limited impact. While their study characterizes configuration sensitivity
across models and workloads, we focus on the load-dependent behavior of the
chunked-prefill token budget within a serving configuration and adapt this
budget online as scheduling pressure changes.

\section{Conclusion}

We presented Prefill-Pressure Adaptive Scheduling (P-PAS), a lightweight
vLLM scheduler policy that adapts the global token budget using
scheduler-visible prefill and decode state. P-PAS retains a large budget when
scheduling pressure is low and switches to a smaller one when prefill--decode
interference increases. Across models, workload shapes, and RTX 5090, A100,
and H100 GPUs, this adaptation maintains low end-to-end latency as load
changes. GPU profiling links the underlying crossover to a trade-off between
large-chunk prefill efficiency and decode progress.

P-PAS+ extends the same principle to heterogeneous request shapes by
incorporating remaining prefill work. On mixed 8k--128k contexts and
16--256-token outputs with Qwen3.8-27B and Nemotron-30B, it improves mean
end-to-end latency by 3.0--3.1\% over the best tested static budget. Overall, scheduling pressure is a useful online control signal:
deployment-specific budget endpoints can be identified with a small static
sweep, and P-PAS switches between them.

\clearpage

\bibliographystyle{plainnat}
\bibliography{references}

\clearpage
\section{Appendix}

\subsection{Two-Endpoint Adaptation}
\label{sec:two_endpoint_analysis}

P-PAS deliberately restricts adaptation to two calibrated budgets,
$B_{\mathrm{cap}}$ and $B_{\max}$, rather than selecting among all possible
MBTs. We quantify the cost of this restriction by comparing the best of the
two endpoints with an oracle that selects the latency-optimal MBT from the
full tested sweep.

For Nemotron-30B with 32k-token prompts and output lengths up to 128 tokens,
the best of the two P-PAS endpoints remains within 1.2\% of the unrestricted
MBT oracle under both low and high scheduling pressure. For longer outputs,
intermediate budgets become more useful; at 256 output tokens under high
pressure, MBT 8k improves latency by about 7\% over the best endpoint.
Thus, the binary controller captures most of the available adaptation benefit
in the short-output regime targeted by P-PAS, but is not universally optimal
for decode-heavy workloads.

\begin{figure}[t]
    \centering

    \begin{subfigure}{\linewidth}
        \centering
        \includegraphics[width=\linewidth]
        {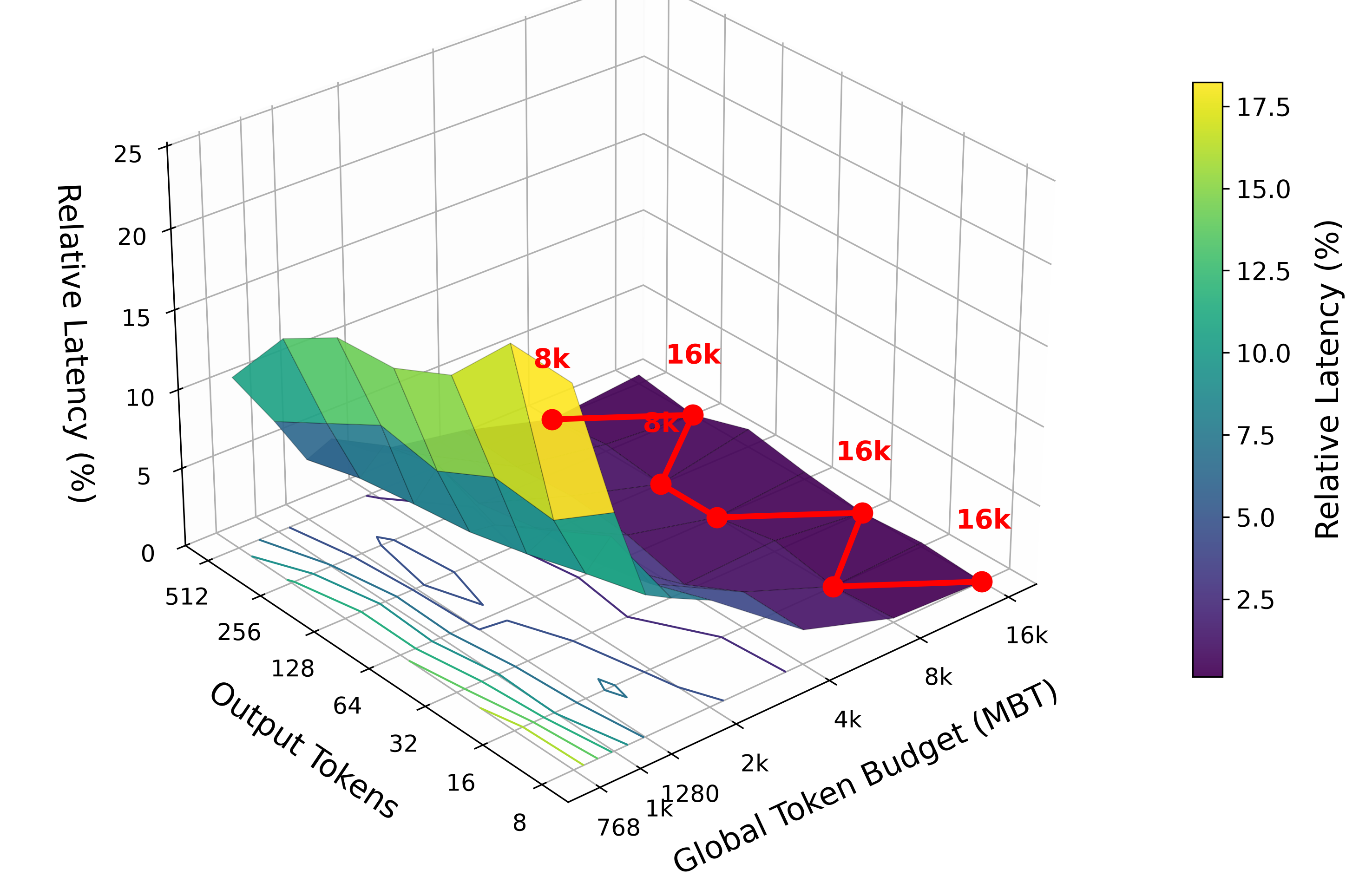}
        \caption{Low pressure: 0.1 requests/s.}
        \label{fig:two_endpoint_low}
    \end{subfigure}

    \vspace{2mm}

    \begin{subfigure}{\linewidth}
        \centering
        \includegraphics[width=\linewidth]
        {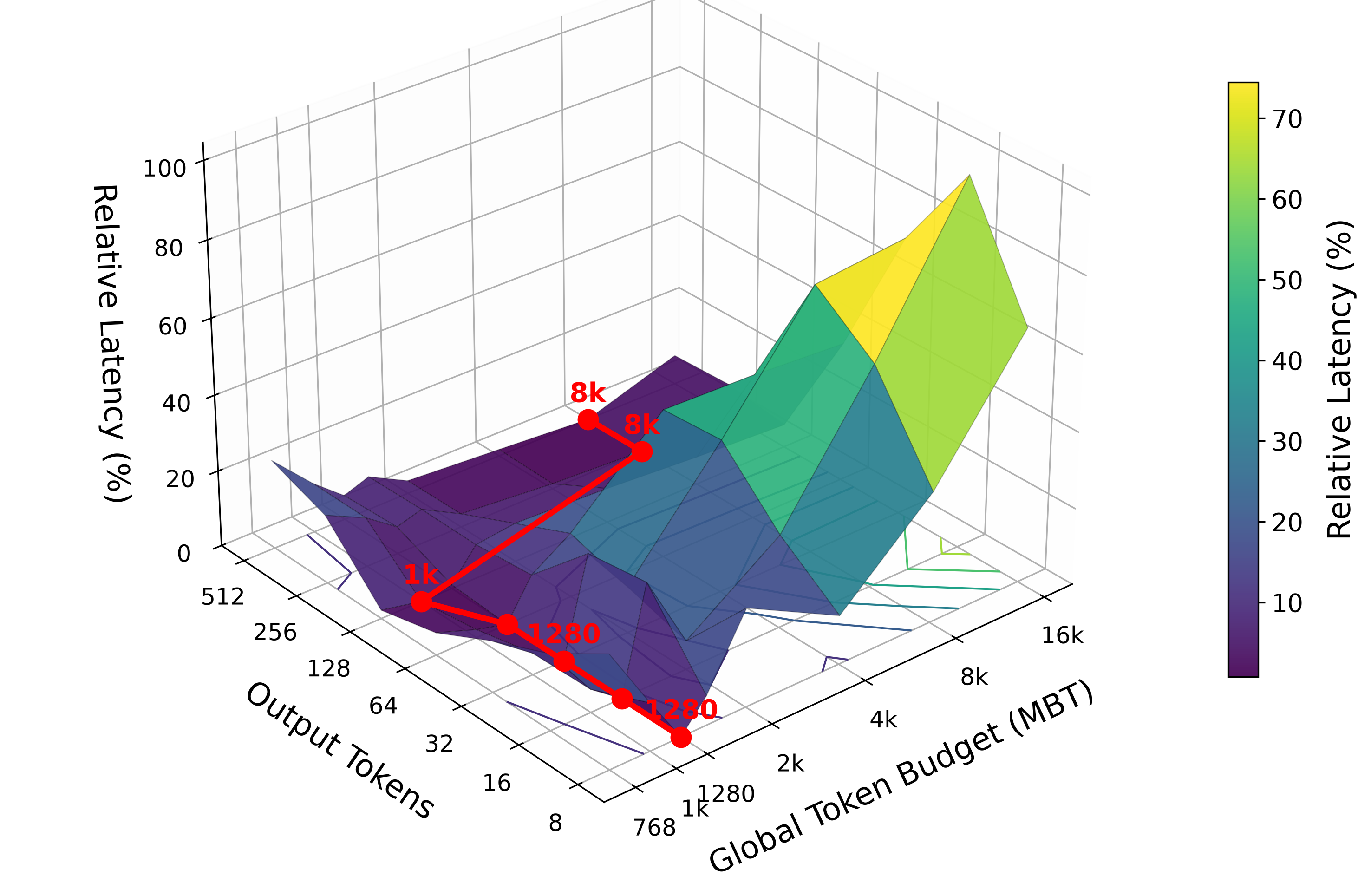}
        \caption{High pressure: 1.0 requests/s.}
        \label{fig:two_endpoint_high}
    \end{subfigure}

    \caption{Relative end-to-end latency across global token budgets for
    NVIDIA Nemotron-3.5-Lightning-30B-A3B-NVFP4 with 32k-token prompts on
    the RTX 5090. Output length varies from 8 to 512 tokens. The red line
    marks the latency-optimal MBT for each output length. Under low pressure,
    large budgets dominate most configurations; under high pressure, smaller
    budgets become preferable for short outputs, while intermediate budgets
    become competitive as output length increases. Measurements use
    deterministic arrivals and one seed.}
    \label{fig:two_endpoint_landscape}
\end{figure}

\subsection{Primary Qwen2.5-3B Results}
\label{app:primary_detailed_results}

Table~\ref{tab:qwen3b_full_results} reports the complete fixed-MBT sweep for
the primary Qwen2.5-3B experiment. The results expose the load-dependent
crossover directly: large budgets are competitive at low pressure, while
smaller budgets become preferable as pressure increases.

\begin{table*}[!t]
\centering
\caption{Detailed results for Qwen2.5-3B on the RTX 5090 with 25k-token
prompts and 32-token outputs. Values are averages over five seeds.
Lower is better for all metrics; the best result for each burst rate and
metric is shown in bold.}
\label{tab:qwen3b_full_results}

\small
\setlength{\tabcolsep}{4pt}

\begin{tabular}{clrrrrr}
\toprule
Burst & Scheduler
& Avg. Lat. [s]
& P95 Lat. [s]
& TTFT [s]
& TPOT [s]
& Makespan [s] \\
\midrule

\multirow{6}{*}{0.4}
& MBT 1k  & 3.503 & 5.393 & 2.379 & 0.0363 & 39.476 \\
& MBT 2k  & 3.348 & 5.168 & 2.079 & 0.0409 & 39.226 \\
& MBT 4k  & 2.993 & 4.980 & 1.850 & 0.0369 & 39.089 \\
& MBT 8k  & 3.285 & 6.165 & 1.985 & 0.0419 & 39.036 \\
& MBT 16k & 2.761 & \textbf{4.502} & 1.695 & 0.0344 & 39.000 \\
& P-PAS   & \textbf{2.735} & 4.672 & \textbf{1.678} &
              \textbf{0.0341} & \textbf{38.970} \\
\midrule

\multirow{6}{*}{0.6}
& MBT 1k  & 4.506 & 7.505 & 3.098 & \textbf{0.0454} & 40.181 \\
& MBT 2k  & 4.399 & 7.057 & 2.551 & 0.0596 & 39.661 \\
& MBT 4k  & 4.252 & 7.767 & 2.208 & 0.0659 & 39.241 \\
& MBT 8k  & 4.005 & 7.396 & 2.142 & 0.0601 & 39.040 \\
& MBT 16k & \textbf{3.678} & 6.854 & 2.039 & 0.0529 &
              \textbf{38.982} \\
& P-PAS   & 3.764 & \textbf{6.659} & \textbf{2.020} & 0.0563 & 39.022 \\
\midrule

\multirow{6}{*}{0.8}
& MBT 1k  & 5.280 & 8.774 & 3.741 & \textbf{0.0496} & 44.540 \\
& MBT 2k  & 5.214 & 8.438 & 3.033 & 0.0703 & 43.889 \\
& MBT 4k  & 5.463 & 9.630 & 2.556 & 0.0938 & 43.429 \\
& MBT 8k  & 5.340 & 10.405 & 2.493 & 0.0918 & 43.225 \\
& MBT 16k & 4.900 & 9.909 & \textbf{2.303} & 0.0838 &
              \textbf{43.180} \\
& P-PAS   & \textbf{4.705} & \textbf{8.387} & 2.579 & 0.0686 & 43.473 \\
\midrule

\multirow{6}{*}{1.0}
& MBT 1k  & 8.110 & 13.891 & 6.300 & \textbf{0.0584} & 49.369 \\
& MBT 2k  & 7.451 & 12.155 & 4.715 & 0.0882 & 46.926 \\
& MBT 4k  & 8.041 & 12.390 & 3.933 & 0.1325 & 45.846 \\
& MBT 8k  & 8.625 & 14.029 & 3.591 & 0.1624 & 45.297 \\
& MBT 16k & 8.420 & 14.668 & \textbf{3.466} & 0.1598 &
              \textbf{45.110} \\
& P-PAS   & \textbf{7.037} & \textbf{11.796} & 4.288 & 0.0887 & 46.326 \\
\midrule

\multirow{6}{*}{1.2}
& MBT 1k  & 12.643 & 21.440 & 10.599 & \textbf{0.0660} & 56.444 \\
& MBT 2k  & 10.687 & 17.237 & 7.382 & 0.1066 & 49.907 \\
& MBT 4k  & 11.206 & \textbf{16.478} & 6.001 & 0.1679 & 47.301 \\
& MBT 8k  & 13.016 & 18.148 & 5.403 & 0.2456 & 46.275 \\
& MBT 16k & 13.264 & 19.493 & \textbf{5.226} & 0.2593 &
              \textbf{45.670} \\
& P-PAS   & \textbf{10.251} & 16.869 & 6.856 & 0.1095 & 48.755 \\
\midrule

\multirow{6}{*}{1.4}
& MBT 1k  & 14.614 & 24.395 & 12.567 & \textbf{0.0660} & 60.081 \\
& MBT 2k  & \textbf{12.560} & 19.881 & 9.180 & 0.1091 & 53.204 \\
& MBT 4k  & 13.050 & \textbf{18.978} & 7.217 & 0.1881 & 49.297 \\
& MBT 8k  & 15.156 & 20.466 & 6.319 & 0.2851 & 47.831 \\
& MBT 16k & 16.649 & 23.004 & \textbf{6.066} & 0.3414 &
              \textbf{47.217} \\
& P-PAS   & 12.895 & 20.537 & 9.416 & 0.1122 & 53.688 \\
\bottomrule
\end{tabular}
\end{table*}


\clearpage

Table~\ref{tab:qwen3b_full_results} shows a small reversal at the highest burst
rate of the main sweep ($\lambda=1.4$), where MBT 2k outperforms P-PAS.
To verify that this does not indicate increasing divergence at higher
pressure, we additionally evaluate both configurations at
$\lambda=1.6$ requests/s.

\begin{table}[H]
\centering
\caption{High-load follow-up for Qwen2.5-3B on the RTX 5090 at
$\lambda=1.6$ requests/s with 25k-token prompts and 32-token outputs.
Values are averages over five seeds.}
\label{tab:high_load_followup}

\small
\setlength{\tabcolsep}{3.5pt}

\begin{tabular}{lrrr}
\toprule
Scheduler & Avg. E2E [s] & P95 E2E [s] & Makespan [s] \\
\midrule
MBT 2k & 14.763 & 22.821 & 56.399 \\
P-PAS  & \textbf{14.324} & \textbf{22.373} & \textbf{55.562} \\
\bottomrule
\end{tabular}
\end{table}

\subsection{Robustness Across Models and GPUs}
\label{app:robustness_detailed}

The following tables provide the absolute values underlying the robustness
results reported in the main paper.

\begin{table}[H]
\centering
\caption{Performance across representative load regimes and model
configurations on the RTX 5090. Results are averages over five seeds.
Lower is better; the best result for each model, burst rate, and metric
is shown in bold.}
\label{tab:rtx5090_load_regimes}

\scriptsize
\setlength{\tabcolsep}{1.8pt}

\begin{tabular}{llrrrrr}
\toprule
Burst & Scheduler & E2E & P95 & TTFT & TPOT & Makespan \\
      &           & [s] & [s] & [s] & [s] & [s] \\
\midrule

\multicolumn{7}{l}{\textbf{Qwen2.5-0.5B}} \\
1.0 & MBT 2k
    & 0.642 & 1.133 & 0.476 & 0.0053 & 40.496 \\
    & MBT 16k
    & \textbf{0.566} & \textbf{0.955} & 0.424 &
      \textbf{0.0046} & 40.430 \\
    & P-PAS
    & 0.571 & 0.973 & \textbf{0.423} & 0.0048 &
      \textbf{40.430} \\
\addlinespace[2pt]

4.0 & MBT 2k
    & 2.655 & 4.656 & 1.936 & \textbf{0.0232} & 43.448 \\
    & MBT 16k
    & 3.529 & 6.160 & \textbf{1.073} & 0.0792 &
      \textbf{43.421} \\
    & P-PAS
    & \textbf{2.576} & \textbf{4.628} & 1.849 & 0.0235 &
      \textbf{43.421} \\
\midrule

\multicolumn{7}{l}{\textbf{Qwen2.5-3B}} \\
0.4 & MBT 2k
    & 3.348 & 5.168 & 2.079 & 0.0409 & 39.226 \\
    & MBT 16k
    & 2.761 & \textbf{4.502} & 1.695 & 0.0344 & 39.000 \\
    & P-PAS
    & \textbf{2.735} & 4.672 & \textbf{1.678} &
      \textbf{0.0341} & \textbf{38.970} \\
\addlinespace[2pt]

1.0 & MBT 2k
    & 7.451 & 12.155 & 4.715 & \textbf{0.0882} & 46.926 \\
    & MBT 16k
    & 8.420 & 14.668 & \textbf{3.466} & 0.1598 &
      \textbf{45.110} \\
    & P-PAS
    & \textbf{7.037} & \textbf{11.796} & 4.288 & 0.0887 & 46.326 \\
\midrule

\multicolumn{7}{l}{\textbf{SmolLM3-3B}} \\
0.4 & MBT 2k
    & 3.658 & 5.638 & 2.194 & 0.0472 & 39.410 \\
    & MBT 16k
    & \textbf{2.958} & \textbf{4.933} & 1.784 &
      \textbf{0.0379} & \textbf{39.062} \\
    & P-PAS
    & 2.974 & 4.953 & \textbf{1.775} & 0.0387 & 39.064 \\
\addlinespace[2pt]

1.0 & MBT 2k
    & 8.415 & 13.513 & 5.407 & 0.0970 & 48.150 \\
    & MBT 16k
    & 9.318 & 15.007 & \textbf{3.757} & 0.1794 &
      \textbf{45.657} \\
    & P-PAS
    & \textbf{7.680} & \textbf{12.754} & 4.723 &
      \textbf{0.0954} & 47.080 \\
\bottomrule
\end{tabular}
\end{table}

\begin{table}[tb]
\centering
\caption{MBT sensitivity on H100 SXM for Qwen2.5-3B with 25k-token prompts
and 32-token outputs. All values are in seconds except TPOT, which is seconds
per output token. Lower is better; the best result for each burst rate and
metric is shown in bold.}
\label{tab:h100}

\scriptsize
\setlength{\tabcolsep}{1.8pt}

\begin{tabular}{llrrrrr}
\toprule
Burst & Scheduler & E2E & P95 & TTFT & TPOT & Makespan \\
\midrule
\multirow{2}{*}{1.0}
& MBT 2k
& 1.393
& 2.605
& 1.049
& 0.0111
& 40.904 \\
& MBT 16k
& \textbf{1.272}
& \textbf{2.397}
& \textbf{0.949}
& \textbf{0.0104}
& \textbf{40.835} \\
\midrule
\multirow{2}{*}{2.0}
& MBT 2k
& \textbf{2.285}
& \textbf{3.957}
& 1.632
& \textbf{0.0211}
& 43.276 \\
& MBT 16k
& 2.448
& 5.016
& \textbf{1.333}
& 0.0360
& \textbf{43.248} \\
\bottomrule
\end{tabular}
\end{table}

\begin{table*}[!t]
\centering
\caption{Detailed A100 SXM results for Qwen2.5-3B and SmolLM3-3B with
25k-token prompts and 32-token outputs. Values are averages over five seeds.
Lower is better for all metrics; the best result for each model, burst rate,
and metric is shown in bold.}
\label{tab:a100_detailed_results}

\small
\setlength{\tabcolsep}{4pt}

\begin{tabular}{llcrrrrr}
\toprule
Model & Burst & Scheduler
& Avg. Lat. [s]
& P95 Lat. [s]
& TTFT [s]
& TPOT [s]
& Makespan [s] \\
\midrule

\multirow{6}{*}{Qwen2.5-3B}
& \multirow{3}{*}{0.4}
& MBT 2k  & 3.179 & 5.037 & 2.049 & 0.0365 & 39.261 \\
&
& MBT 16k & 2.750 & 4.882 & 2.024 & \textbf{0.0234} &
             \textbf{39.096} \\
&
& P-PAS   & \textbf{2.706} & \textbf{4.233} & \textbf{1.818} &
             0.0287 & 39.106 \\
\cmidrule(lr){2-8}

& \multirow{3}{*}{1.0}
& MBT 2k  & 6.853 & \textbf{11.189} & 4.380 &
             \textbf{0.0798} & 46.417 \\
&
& MBT 16k & 7.724 & 13.756 & \textbf{3.384} & 0.1400 &
             \textbf{44.791} \\
&
& P-PAS   & \textbf{6.850} & 11.514 & 4.200 & 0.0855 & 46.190 \\
\midrule

\multirow{6}{*}{SmolLM3-3B}
& \multirow{3}{*}{0.4}
& MBT 2k  & 3.368 & 5.504 & 2.140 & 0.0396 & 39.344 \\
&
& MBT 16k & \textbf{2.764} & \textbf{4.301} & 1.845 &
             \textbf{0.0296} & \textbf{39.139} \\
&
& P-PAS   & 2.798 & 4.400 & \textbf{1.841} & 0.0309 & 39.141 \\
\cmidrule(lr){2-8}

& \multirow{3}{*}{1.0}
& MBT 2k  & 7.244 & \textbf{11.784} & 4.646 &
             \textbf{0.0838} & 46.790 \\
&
& MBT 16k & 7.959 & 14.040 & \textbf{3.453} & 0.1453 &
             \textbf{45.005} \\
&
& P-PAS   & \textbf{7.030} & \textbf{11.784} & 4.315 &
             0.0876 & 46.350 \\
\bottomrule
\end{tabular}
\end{table*}

\FloatBarrier

\subsection{Additional Homogeneous Large-Model Results}
\label{app:large_model_homogeneous}

Tables~\ref{tab:qwen27b_per_condition} and
\ref{tab:nemotron30b_per_condition} report mean $\pm$ standard deviation
across five seeds. Lower is better, and the best result within each burst
rate and metric is shown in bold.

\begin{table}[H]
\centering
\caption{Qwen3.8-27B NVFP4 with 20k-token prompts and 32-token outputs.}
\label{tab:qwen27b_per_condition}

\scriptsize
\setlength{\tabcolsep}{2.2pt}
\renewcommand{\arraystretch}{1.02}

\begin{tabular}{clccc}
\toprule
Burst & Scheduler & Mean E2E [s] & P95 E2E [s] & Makespan [s] \\
\midrule

\multirow{3}{*}{0.2}
& MBT 768
& 4.559 $\pm$ 0.703
& 6.174 $\pm$ 1.344
& 41.359 $\pm$ 10.484 \\
& MBT 2048
& 4.555 $\pm$ 0.829
& 6.712 $\pm$ 1.548
& \textbf{41.019 $\pm$ 10.329} \\
& P-PAS
& \textbf{4.425 $\pm$ 0.817}
& \textbf{6.059 $\pm$ 1.289}
& 41.099 $\pm$ 10.407 \\
\midrule

\multirow{3}{*}{0.4}
& MBT 768
& \textbf{6.387 $\pm$ 0.559}
& \textbf{9.453 $\pm$ 1.686}
& 38.895 $\pm$ 6.672 \\
& MBT 2048
& 7.245 $\pm$ 0.700
& 10.618 $\pm$ 0.787
& \textbf{38.287 $\pm$ 6.372} \\
& P-PAS
& 6.504 $\pm$ 0.471
& 10.532 $\pm$ 1.371
& 38.486 $\pm$ 6.559 \\
\midrule

\multirow{3}{*}{0.6}
& MBT 768
& 9.521 $\pm$ 1.574
& \textbf{13.602 $\pm$ 1.789}
& 43.437 $\pm$ 8.529 \\
& MBT 2048
& 11.401 $\pm$ 1.988
& 15.671 $\pm$ 2.159
& \textbf{41.106 $\pm$ 8.167} \\
& P-PAS
& \textbf{9.475 $\pm$ 1.452}
& 13.901 $\pm$ 1.711
& 42.982 $\pm$ 8.608 \\
\midrule

\multirow{3}{*}{0.8}
& MBT 768
& 12.193 $\pm$ 1.291
& 17.423 $\pm$ 2.124
& 47.167 $\pm$ 9.383 \\
& MBT 2048
& 13.725 $\pm$ 1.051
& 19.820 $\pm$ 1.791
& \textbf{44.070 $\pm$ 8.091} \\
& P-PAS
& \textbf{11.714 $\pm$ 1.042}
& \textbf{16.933 $\pm$ 1.677}
& 46.534 $\pm$ 9.020 \\
\midrule

\multirow{3}{*}{1.0}
& MBT 768
& 15.185 $\pm$ 2.355
& \textbf{22.966 $\pm$ 4.200}
& 56.918 $\pm$ 11.537 \\
& MBT 2048
& 16.792 $\pm$ 2.404
& 24.153 $\pm$ 2.586
& \textbf{53.013 $\pm$ 10.956} \\
& P-PAS
& \textbf{15.099 $\pm$ 2.375}
& 23.028 $\pm$ 4.296
& 56.682 $\pm$ 11.605 \\
\bottomrule
\end{tabular}
\end{table}

\begin{table}[tb]
\centering
\caption{NVIDIA Nemotron-3.5-Lightning-30B-A3B-NVFP4 with 20k-token
prompts and 32-token outputs.}
\label{tab:nemotron30b_per_condition}

\scriptsize
\setlength{\tabcolsep}{2.2pt}
\renewcommand{\arraystretch}{1.02}

\begin{tabular}{clccc}
\toprule
Burst & Scheduler & Mean E2E [s] & P95 E2E [s] & Makespan [s] \\
\midrule

\multirow{4}{*}{0.5}
& MBT 1280
& 1.556 $\pm$ 0.234
& 2.607 $\pm$ 0.442
& 34.619 $\pm$ 6.131 \\
& MBT 2048
& 1.507 $\pm$ 0.211
& 2.544 $\pm$ 0.461
& 34.566 $\pm$ 6.167 \\
& MBT 16384
& \textbf{1.440 $\pm$ 0.193}
& 2.475 $\pm$ 0.474
& \textbf{34.528 $\pm$ 6.165} \\
& P-PAS
& 1.444 $\pm$ 0.195
& \textbf{2.458 $\pm$ 0.443}
& 34.529 $\pm$ 6.167 \\
\midrule

\multirow{4}{*}{1.0}
& MBT 1280
& \textbf{2.296 $\pm$ 0.232}
& \textbf{3.774 $\pm$ 0.366}
& 40.423 $\pm$ 5.480 \\
& MBT 2048
& 2.307 $\pm$ 0.264
& 4.237 $\pm$ 0.145
& 40.364 $\pm$ 5.470 \\
& MBT 16384
& 2.566 $\pm$ 0.455
& 5.482 $\pm$ 1.056
& \textbf{40.325 $\pm$ 5.477} \\
& P-PAS
& 2.333 $\pm$ 0.294
& 4.368 $\pm$ 0.294
& 40.338 $\pm$ 5.480 \\
\midrule

\multirow{4}{*}{1.5}
& MBT 1280
& 4.462 $\pm$ 1.272
& \textbf{7.167 $\pm$ 2.175}
& 39.737 $\pm$ 5.444 \\
& MBT 2048
& 4.740 $\pm$ 1.332
& 7.430 $\pm$ 2.060
& 39.383 $\pm$ 5.427 \\
& MBT 16384
& 6.869 $\pm$ 2.597
& 11.014 $\pm$ 2.491
& \textbf{39.108 $\pm$ 5.480} \\
& P-PAS
& \textbf{4.449 $\pm$ 1.202}
& 7.498 $\pm$ 1.815
& 39.614 $\pm$ 5.458 \\
\midrule

\multirow{4}{*}{2.0}
& MBT 1280
& 5.663 $\pm$ 0.889
& 8.362 $\pm$ 1.326
& 40.892 $\pm$ 5.802 \\
& MBT 2048
& 6.012 $\pm$ 0.877
& 8.478 $\pm$ 1.326
& 40.347 $\pm$ 5.832 \\
& MBT 16384
& 9.971 $\pm$ 1.345
& 14.640 $\pm$ 1.238
& \textbf{39.848 $\pm$ 5.874} \\
& P-PAS
& \textbf{5.588 $\pm$ 0.897}
& \textbf{8.236 $\pm$ 1.355}
& 40.753 $\pm$ 5.833 \\
\bottomrule
\end{tabular}
\end{table}

\subsection{Heterogeneous P-PAS+ Workload Definition}
\label{app:heterogeneous_schedules}

We evaluate six dynamic conditions per model, combining two burst rates
with three burst durations; each burst is preceded and followed by a
low-rate phase.

For Qwen, $\lambda_b\in\{0.1,0.4\}$ requests/s, with prompt lengths
$\{8\mathrm{k},16\mathrm{k},32\mathrm{k},64\mathrm{k}\}$. For Nemotron,
$\lambda_b\in\{0.2,0.8\}$ requests/s, with prompt lengths
$\{16\mathrm{k},32\mathrm{k},64\mathrm{k},128\mathrm{k}\}$. Both use
output lengths $\{16,32,64,128,256\}$.

Each condition is evaluated with 11 seeds (0--10). For each seed, the same
request trace is replayed across all schedulers, enabling exact request-level
matching. Let $M_{s,w}$ denote the mean end-to-end latency of scheduler $s$
for workload condition $w$, aggregated over the repeated runs. 
We report improvement over a baseline as

\begin{equation}
\mathrm{Improvement}
=
\left(
1 -
\frac{1}{N}
\sum_{w=1}^{N}
\frac{M_{\mathrm{P\text{-}PAS+},w}}
     {M_{\mathrm{baseline},w}}
\right)
\times 100\%,
\end{equation}
where $N=6$, weighting all workload conditions equally.

\begin{table}[tb]
\centering
\caption{Dynamic heterogeneous workload schedules.}
\label{tab:heterogeneous_schedules}

\small
\setlength{\tabcolsep}{3pt}
\renewcommand{\arraystretch}{0.95}

\begin{tabular}{lcc}
\toprule
Model & Burst duration & Phase schedule [requests/s] \\
\midrule

\multirow{3}{*}{Qwen3.8-27B}
& 10 s & $20{:}0.05,\ 10{:}\lambda_b,\ 20{:}0.05$ \\
& 20 s & $20{:}0.05,\ 20{:}\lambda_b,\ 20{:}0.05$ \\
& 40 s & $20{:}0.05,\ 40{:}\lambda_b,\ 20{:}0.05$ \\

\multirow{3}{*}{Nemotron-30B}
& 10 s & $20{:}0.1,\ 10{:}\lambda_b,\ 20{:}0.1$ \\
& 20 s & $20{:}0.1,\ 20{:}\lambda_b,\ 20{:}0.1$ \\
& 40 s & $20{:}0.1,\ 40{:}\lambda_b,\ 20{:}0.1$ \\
\bottomrule
\end{tabular}
\end{table}

\end{document}